\documentstyle[11pt,amsmath,amssymb,epsf,epic,cite, xcolor]{article}
\newcommand{\un}{{\mathbb I}}
\newcommand{\ra}{\rightarrow}
\newcommand{\tr}{{\rm tr}}
\newcommand{\spa}{{\rm span}}
\newcommand{\ran}{{\rm Ran\, }}
\renewcommand{\ker}{{\rm Ker\, }}
\newcommand{\bra}{\langle}
\newcommand{\ket}{\rangle}
\newcommand{\be}{\begin{equation}}
\newcommand{\ee}{\end{equation}}
\newcommand{\bea}{\begin{eqnarray}}
\newcommand{\eea}{\end{eqnarray}}
\newcommand{\eps}{\epsilon}

\newcommand{\ffi}{\varphi}

\newcommand{\ode}{{\cal O}}
\newcommand{\grintl}{[\kern-.18em [}
\newcommand{\grintr}{]\kern-.18em ]}

\newcounter{resultcounter}[section]

\newtheorem{thm}[resultcounter]{Theorem}
\newtheorem{lem}[resultcounter]{Lemma}
\newtheorem{prop}[resultcounter]{Proposition}
\newtheorem{cor}[resultcounter]{Corollary}
\newtheorem{definition}[resultcounter]{Definition}
\newtheorem{rem}[resultcounter]{Remark}
\numberwithin{equation}{section}
\def\bed{\begin{definition}}
\def\eed{\end{definition}}

 \def\cB{{\cal B}} 
\def\cD{{\cal D}}  
 \def\cH{{\cal H}} 
  \def\cL{{\cal L}}
  
  \def\cR{{\cal R}}

\newcommand{\R}{{\mathbb R}}
\newcommand{\N}{{\mathbb N}}

\newcommand{\C}{{\mathbb C}}

\renewcommand{\P}{{\mathbb P}}

\def\qed{\hfill $\Box$\medskip}
\newcommand{\qrm}{QRM }
\newcommand{\diag}{{\rm Diag}}
\newcommand{\offdiag}{{\rm Offdiag}}

\newcommand{\rank}{{\rm Rank\,}}

\newcommand{\gh}[1]{{\color[rgb]{0,0,1}{#1}}}
\newcommand{\aj}[1]{{\color[rgb]{0,0,0}{#1}}}

\begin{document}

\title{Perturbation Analysis of Quantum Reset Models}

\author{G\'eraldine Haack\footnote{Department of Applied Physics, University of Geneva, Chemin de Pinchat 22, 1227 Carouge, Gen\`eve, Switzerland} \ $\&$ Alain Joye \footnote{ Univ. Grenoble Alpes, CNRS, Institut Fourier, F-38000 Grenoble, France}}

\date{ }

\maketitle

\abstract{This paper is devoted to the analysis of Lindblad operators \aj{of Quantum Reset Models}, describing the effective dynamics of tri-partite quantum systems \aj{subject to stochastic resets.} We consider a chain of three independent subsystems, coupled by a Hamiltonian term. The two subsystems at each end of the chain are driven, independently from each other, by a reset Lindbladian, while the center system is driven by a Hamiltonian. Under generic assumptions on the coupling term, we prove the existence of a unique steady state for the perturbed reset Lindbladian, analytic in the coupling constant. We further analyze the large times dynamics of the corresponding CPTP Markov semigroup that describes the approach to the steady state. We illustrate these results with concrete exemples corresponding to realistic open quantum systems.}

\tableofcontents

\section{Introduction}

\aj{
A major challenge when investigating small quantum systems is to assess their dynamics when coupled to several environments that put the system in an out-of-equilibrium situation. To do so, one often resorts to effective master equations governing the reduced density operator for the small system.
Under the Born-Markov approximation (that involves weak system-bath coupling and short bath time-correlations), the evolution equation for the reduced density operator becomes linear, and is cast into the form of a Lindblad-type master equation \cite{GKLS} for the corresponding map to be CPTP (Completely Positive and Trace Preserving). A Hamiltonian approach using perturbation theory is probably the most standard way to derive such a (continuous in time) effective evolution equation for the reduced quantum system \cite{Breuer, Schaller}. For an account of mathematical results, we refer the reader to the review \cite{DerFru}. 
Alternatively, repeated-interaction schemes (discrete in time) have attracted lots of attention among both mathematicians \cite{KumMaa, AttalPautrat, BJM0, BruPil, BJMRev, HJPR} and physicists \cite{Rau63, Barra15, Barra17, Lorenzo17, Pezzuto16, Strasberg17, Seah19}, especially in the context of quantum thermodynamics. Exact solutions for the asymptotic steady states generated by both types of dynamics can in general be derived for quantum systems with low dimensional Hilbert space only.

Appealing master equations to investigate the dynamics of higher dimensional quantum systems are provided by a specific class of models, known as Quantum Reset Models (QRM hereafter). These models can be viewed as a natural extension of classical stochastic models, see \cite{Kampen} for a review and \cite{Evans11} for an example treating diffusion processes. Remarkably, QRM can be formulated in terms of Lindblad master equations so that they generate CPTP maps. This is achieved by making specific choices of dissipation channels (corresponding to a fully depolarized quantum channel), see \cite{Hartmann17, MSM18, Tavakoli18} for examples in specific physical setups. These QRM, thanks to their structural simplicity, present the strong advantage to allow for analytical solutions for the reduced density operator of multipartite quantum systems and have been successfully exploited to assess the dynamics of specific quantum systems, namely small quantum thermal machines made of a few qubits, qutrits or higher-dimensional quantum systems \cite{LPS, Skrzypczyk11, Brunner14, Brask15, Tavakoli18, Tavakoli20}. 

In this work, we 
raise the question to which extent general properties of the dynamics generated by QRM can be analyzed mathematically. Our aim is to go beyond specific models to determine generic properties of the dynamics of QRM, {\it i.e.} induced by the mathematical structure itself of the QRM. A first step in that direction is performed in the recent work  \cite{Rose18} where a single system driven by a Lindbladian subject to a reset process is considered. The spectral properties of the total Lindbladian perturbed by the reset processes are established, under the assumption that the unperturbed Lindbadian possesses a unique stationary state. In the present paper, we consider QRM describing the dynamics of more complex structures that are therefore intrinsically degenerate and not amenable to the cases dealt with above. We reach a two-fold objective. On the one hand, we show that those degenerate QRM nevertheless allow for a complete mathematical treatment revealing a rich structure. On the other hand, we demonstrate the relevance of our perturbative analysis to assess the dynamics of realistic multipartite quantum systems characterized by Hilbert spaces of dimension as high as 8.}

More precisely, \aj{our generic model is made of } a tripartite structure, $A-C-B$, where $A$ and $B$ are the two quantum systems subject to reset processes, and $C$ is a central system with its own free evolution. The three subsystems are weakly interacting through a Hamiltonian. %This model is for instance suitable to describe a chain of small quantum systems, with each end coupled to its own reservoir.
We first \aj{recall} that \qrm are always characterised by Lindblad generators, \aj{with explicit dissipators}. Then we analyse the spectral properties of the resulting Lindbladians and the dynamics of the tri-partite system they generate, under generic hypotheses on the coupling term. We conduct this analysis first in absence of interaction between the $A-C-B$ parts of the Hilbert space they are defined on, which gives rise to an uncoupled Lindbladian displaying large degeneracies, {\it i.e.} a large subspace of invariant states. Then, we introduce a generic interaction between these different parts and perform a perturbative analysis in the coupling constant. We prove uniqueness of an invariant steady state under the coupled dynamics, analytic in the coupling constant, and provide a description of the converging power series of this non-equilibrium steady state that develops in the small system. Building up on our spectral analysis, we elucidate the long time properties of the dynamics of the tri-partite system and its approach to the steady state. Finally, we focus on the case where the uncoupled system has no Hamiltonian drive and we describe in particular the emergence of a natural classical Markov process in the description of the large time behaviour of the coupled system. The paper closes with the study of two examples illustrating the key features of this analysis: the systems A and B are two qubits while the central system C is of arbitrary dimension $N$ and the uncoupled dynamics has no Hamiltonian drive. For a rather general choice of QRM coupled dynamics, we compute the leading order of the steady state for $N$ arbitrary and, for $N=2$ -- when C is another qubit -- we determine the steady state up to order three in the coupling constant as well as the associated classical Markov process.

\section{Mathematical framework}

\subsection{Simple Hilbert space setup}\label{basics}

As a warmup, we consider a single quantum system of finite dimension characterized by its Hamiltonian $H$ defined on its Hilbert space $\cH$ which is coupled to $M$ reservoirs. 
QRMs assume the state of the quantum system to be reset to a given state $\tau_l$ with probablity $\gamma_l \, dt$ within each time interval $dt$. The QRM-type evolution equation is given by \aj{\cite{LPS, Brask15, Hartmann17}}:
\begin{align}\label{simpleqrm}
\dot{\rho}(t)=-i [H, \rho]+\sum_{l=1}^M\gamma_l(\tau_l \, \tr(\rho)-\rho)\,.
\end{align}
The operator $\rho$ is the reduced density operator of the system defined on $\cH$, and $\gamma_l$ characterizes the coupling rate to the reservoir $l$, $l= 1, \ldots, M$.
\medskip

For the sake of comparison with our main concern --tri-partite systems-- and to set the notation, we discuss the dynamics of \qrm defined in this simple setup, \aj{essentially along the lines of \cite{Rose18}}. We provide a full description of its generic features, under the following assumptions. \\

\noindent
{\bf Gen}:\\
Let   $\cH$ be a Hilbert space, with $\dim \cH=N<\infty$. The dissipative part of the generator is characterised by  
\begin{itemize}
\item $\{\tau_l\}_{1\leq l \leq M}$ a collection of density matrices on $\cH$, {\it i.e.} $\tau_l\in \cB(\cH)$, with $\tau_l\geq 0$ and $\tr (\tau_l)=1$, for all $l\in {1, \dots, M}$, 
\item $\gamma_l >0$,  $l\in {1, \dots, M}$, the collection of associated non\gh{-}zero rates for the coupling to the $M$ baths.
\end{itemize}
The Hamiltonian part of the generator, $H=H^*\in\cB(\cH)$, is generic in the spectral sense 
\begin{itemize}
\item $\sigma(H)=\{e_1, e_2, \dots, e_N\}$, consists of simple eigenvalues with associated normalised eigenvectors denoted by $\{\ffi_j\}_{1\leq j\leq N}$, {\it i.e.} $H\ffi_j = e_j\ffi_j$, $j\in\{1, \cdots, N\}$,
\item The differences (Bohr frequencies) $\{e_j-e_k\}_{j\neq k}$ are all distinct.
\end{itemize}
The generator of \qrm is thus the (super-)operator $\cL\in\cB(\cB(\cH))$ defined by 
\begin{align}\label{simpleqrm}
\cL (\rho)=-i [H, \rho]+\sum_{l=1}^M\gamma_l(\tau_l \, \tr(\rho)-\rho),
\end{align}
where $\rho$ here is arbitrary in $\cB(\cH)$, such that the dynamics of the \qrm reads
\begin{align}\label{dynqrm}
\dot{\rho}(t)=\cL (\rho(t)), \ \ t\in (0,\infty), \ \ \rho(0)=\rho_0\in \cB(\cH).
\end{align}
In case $\rho\in {\cal DM(H)}$, the set of density matrices 
${\cal DM(H)}=\{\rho\in \cB(\cH) \, | \, \rho \geq 0, \tr(\rho)=1\}$, the trace factor in (\ref{simpleqrm}) disappears. Indeed, we will see below in a more general framework that the operator $\cL$ enjoys further symmetries, being a Lindblad operator, see Proposition \ref{lind}; in particular if $\rho_0\in {\cal DM(H)}$, $\rho(t)\in {\cal DM(H)}$, for all $t\in [0,\infty)$. 

However, we perform the full spectral analysis of $\cL$ as an operator on $\cB(\cH)$ and, accordingly, solve the equation (\ref{dynqrm}) without resorting to these symmetries. 

\medskip

We first combine the density matrices $\tau_l$ with corresponding rates $\gamma_l$ into a single density matrix $T$ with corresponding rate $\Gamma$:
Setting  
\be
\Gamma=\sum_{l=1}^M\gamma_l>0,  \ \ T=\frac{1}{\Gamma }{\sum_{l=1}^M} \gamma_l \tau_l\in {\cal DM(H)},
\ee
 we get that (\ref{simpleqrm}) writes
 \be\label{renl}
 \cL(\rho)=-i[H,\rho]+\Gamma (T\tr (\rho)-\rho).
 \ee

In the sequel, we denote the matrix elements of  any $A\in \cB(\cH)$ in the basis $\{\ffi_j\}_{1\leq j\leq N}$ by $A_{jk}=\bra \ffi_j | A \ffi_k\ket$, and the operator $|\ffi\ket\bra\psi |\in \cB(\cH)$, for $\ffi, \psi\in \cH$, is defined by $|\ffi\ket\bra\psi | : \eta\mapsto \ffi \bra \psi|\eta\ket$.

\begin{lem}
Under our assumptions {\bf Gen}, the operator $\cL: \cB(\cH)\ra \cB(\cH)$ {  defined by (\ref{renl}) } is diagonalisable with spectrum given by
\begin{align}
\sigma(\cL)=\{0, -\Gamma\}\cup \{-i(e_j-e_k)-\Gamma\}_{j\neq k}\}.
\end{align}
All eigenvalues are simple, except $-\Gamma$ which has multiplicity $N-1$.\\
Moreover, the solution to (\ref{dynqrm}) reads
{  
\begin{align}\label{basisindep}
\rho(t)=e^{-t(i[H,\cdot]+\Gamma)}\big(\rho_0-{\em \tr}(\rho_0)\Gamma \big(i[H,\cdot]+\Gamma\big)^{-1}(T) \big)+{\em \tr}(\rho_0)\Gamma \big(i[H,\cdot]+\Gamma\big)^{-1}(T).
\end{align}
}
Expressed in the eigenbasis of $H$, this means that, with $\lambda_{jk}=i(e_j-e_k)+\Gamma$,
{  
\begin{align}\label{matrixelmt}
\rho_{jk}(t)=e^{-t\lambda_{jk}} {\rho_0}_{jk}+ {\em \tr}(\rho_0)\Gamma\frac{T_{jk}}{\lambda_{jk}}\big(1-e^{-t\lambda_{jk}} \big), \ \ \mbox{for all } \ 1\leq j,k\leq N.
\end{align}
}
\end{lem}
\begin{rem}
i) In the limit $t\ra\infty$ the steady state is independent of the initial condition and reads
{  
\begin{align}
\rho^{SS} \equiv \lim_{t\ra \infty}\rho(t)=\Gamma \big(i[H,\cdot]+\Gamma\big)^{-1}(T)
\end{align}
}
ii) In particular, for $\rho_0\in {\cal DM(H)}$, all populations decay to { $T_{jj}$ } at the same exponential rate without oscillations 
{  
$\rho_{jj}(t)=e^{-t\Gamma} {\rho_0}_{jj}+ T_{jj}\big(1-e^{-t\Gamma}\big)$.
}\\
\aj{iii)
The result is known, see {\it e.g.} \cite{Rose18}; we provide a proof for the sake of comparison with those of the sections to come.}
\end{rem}
\begin{proof}
{ We first deal with  the dynamical aspects and note that 
$\cL(\cdot) =-(i[H,\cdot ]+\Gamma \cdot)+\Gamma T\tr (\cdot)$ , with $\tr\ T=1$ 
}
implies {  $\tr \cL(\rho)=0$} for any $\rho\in\cB(\cH)$, so that the trace is conserved by (\ref{dynqrm}). Hence, considering the $jk$ matrix element of the differential equation (\ref{dynqrm}) we get
{  
 \begin{align}
 \dot\rho_{jk}=-\lambda_{jk}\rho_{jk}+\Gamma T_{jk}\tr(\rho_0) \ \ \mbox{where} \ \ \lambda_{jk}\neq 0,
\end{align}
}
which yields (\ref{matrixelmt}). The basis independent formulation (\ref{basisindep}) follows by the decomposition $\rho=\sum_{1\leq j,k\leq N}\rho_{jk}|\ffi_j\ket\bra \ffi_k|$ and the observation  
\begin{align}\label{obs}
i[H,|\ffi_j\ket\bra \ffi_k|]+\Gamma|\ffi_j\ket\bra \ffi_k|=\lambda_{jk}|\ffi_j\ket\bra \ffi_k|,
\end{align}
{  
which yields  ${\big(i[H,\cdot]+\Gamma\big)^{-1}(T)}_{jk}= T_{jk}/\lambda_{jk}$.\\
}

On the spectral side, the observation above immediately yields 
{  
$ \cL(|\ffi_j\ket\bra \ffi_k|)=-\lambda_{jk}|\ffi_j\ket\bra \ffi_k|$ for $j\neq k$, showing  $\{-\lambda_{jk}\}_{j\neq k}$ are simple eigenvalues by our genericity assumption. 
To compute the other nonzero eigenvalues of $\cL$, we note that  if $\rho$ is an eigenvector of $\cL$ associated with an eigenvalue $\lambda$, then $\lambda \tr \rho =0$.
Hence $\lambda\neq 0$ implies $\tr \rho=0$.  Thus, considering the $N-1$ dimensional  subspace of diagonal traceless matrices in the eigenbasis of $H$,
$\{\rho = \sum_{1\leq j\leq N} r_j |\ffi_j\ket\bra \ffi_j| \ \, | \, \, \sum_{1\leq j\leq N}r_j=0\}$, and making use of the identity  $\cL(|\ffi_j\ket\bra \ffi_j|)= \Gamma (T - |\ffi_j\ket\bra \ffi_j|) $, for any $j$,
we see that it coincides with  $\ker (\cL +\Gamma \un)$.
 Finally, the one-dimensional kernel of $\cL$ is spanned by 
$\Gamma \big(i[H,\cdot]+\Gamma\big)^{-1}(T)$: the inverse is well defined thanks to (\ref{obs}), it has matrix elements $\Gamma T_{jk}/\lambda_{jk}$, and trace one. Thus 
\be
\cL(\Gamma \big(i[H,\cdot]+\Gamma\big)^{-1}(T))=-\Gamma \big(i[H,\cdot]+\Gamma \cdot \big)(\big(i[H,\cdot]+\Gamma\big)^{-1}(T))+\Gamma T =0.
\ee
}
 \qed
\end{proof}

\subsection{Tri-partite Hilbert spaces}\label{tripartite}

\medskip

We define here the tri-partite systems whose dynamical properties are studied in this paper.

Consider $\cH=\cH_A\otimes\cH_C\otimes\cH_B$, where $\cH_\#$ are Hilbert spaces, with dimensions noted $n_\#<\infty$, where $\#\in\{A,B,C\}$. Let $\tau_A\in {\cal DM(H}_A)$, $\tau_B\in {\cal DM(H}_B)$ be two density matrices on their respective Hilbert space and $\gamma_A, \gamma_B>0$ two positive rates. Consider three Hamiltonians $H_A, H_B, H_C$ on their respective Hilbert space that further satisfy 
\be\label{comth}
[H_A,\tau_A]=0, \ \ \mbox{ and } \ \  [H_B,\tau_B]=0, 
\ee
while $H_C$ is arbitrary at this point. In applications, the reset state $\tau_\#$ will typically be defined as a Gibbs state at some inverse temperature $\beta_\#$ associated to $H_\#$; {\it i.e.} $\tau_\#=e^{-\beta_\#H_\#}/Z_\#$ which satisfies (\ref{comth}), where $Z_\#$ is the corresponding partition function. In Sec.\ref{sec:2}, we perform the analysis of the uncoupled case (system $A-C-B$ is non-interacting), and in Sec.\ref{sec:3}, we make use of analytic perturbation theory to treat the case where a weak interaction is added to the system $A-C-B$.

\section{The non-interacting tripartite \qrm}
\label{sec:2}

We define the {\it uncoupled} \qrm by the generator
\begin{align}\label{triqrm}
\cL(\rho)=&-i[H_A\otimes \un_{C}\otimes \un_{B}+   \un_{A}\otimes H_C\otimes \un_{B}  + \un_{A}\otimes \un_{C}\otimes H_B ,\rho]\\ \nonumber
&+\gamma_A(\tau_A\otimes \tr_A(\rho)-\rho)+\gamma_B(\tr_B(\rho)\otimes \tau_B-\rho),
\end{align}
where $\un_{\#}$ denotes the identity operator on $\cH_\#$
and $\tr_\#$ denotes the operator on the tensor product of Hilbert spaces with indices different from $\#$, obtained by taking the partial trace over $\cH_\#$. For later purposes, 
$\tr_{\#\#'}$ denotes the operator on the Hilbert space with index different from $\#$ and $\#'$ obtained by taking the partial trace over $\cH_\#\otimes\cH_{\#'}$. For example,
\be
\tr_A : \cB(\cH_A\otimes\cH_C\otimes\cH_B)\ra \cB(\cH_C\otimes\cH_B),\ \  \tr_{AB} : \cB(\cH_A\otimes\cH_C\otimes\cH_B)\ra \cB(\cH_C)
\ee
will be viewed as linear maps.
We shall abuse notations and write $H_\#$ for the Hamiltonian both on $\cH_\#$ and $\cH$, the context making it clear what we mean. Also, we shall denote the non-Hamiltonian part of the generator by 
\be\label{dissip}
\cD(\rho)=\gamma_A(\tau_A\otimes \tr_A(\rho)-\rho)+\gamma_B(\tr_B(\rho)\otimes \tau_B-\rho),
\ee
so that 
$
\cL(\rho)=-i\big[H_A+H_C  +H_B , \rho\big]+\cD(\rho).
$
\begin{rem}
If $n_B=1$, $\cH_B\simeq \C$ and the last tensor product is trivial. Hence the \qrm reduces to $\cL(\rho)=-i\big[H_A+H_C, \rho\big]+\gamma_A(\tau_A\otimes \tr_A(\rho)-\rho)$ on $\cH_A\otimes \cH_C$, while keeping $\gamma_B>0$.
\end{rem}
\medskip

Let us start by a structural result saying that the \qrm at time $t$, $e^{t\cL}(\rho_0)$, with $\rho_0$ a state, is a CPTP map, by \aj{recalling that its generator can be cast under the form of a Lindblad operator, see e.g. \cite{Brask15, Hartmann17, LPS}}. 
More precisely,  the non-Hamiltonian part of their generator 
 (\ref{triqrm}) takes the form of a dissipator, {\it i.e.}
\begin{align}
\sum_jA_j\rho A_j^*-\frac12\{A_j^*A_j,\rho\}=\sum_j \frac12\Big\{[A_j\rho,A_j^*]+[A_j,\rho A_j^*]\Big\}, \ \ \mbox{for } \ \ A_j\in \cB(\cH).
\end{align}
Given (\ref{triqrm}), it is enough to consider $\tau_A\otimes \tr_A(\rho)-\rho$ defined on $\cH=\cH_A\otimes\cH_C$.
\begin{prop}\label{lind} Let $\tau_A=\sum_{k}t_k|\ffi_k\ket\bra\ffi_k|$ be the spectral decomposition of $\tau_A$, where $\{\ffi_k\}_{k}$ is a complete orthonormal basis of $\cH_A$. Then
\begin{align}
\tau_A\otimes {\em \tr}_A(\rho)-\rho=\sum_{j,k}\big(A_{jk}\rho A_{jk}^*-\frac12\{A_{jk}^*A_{jk},\rho\}\big), \ \ \mbox{where} \ \ A_{jk}=\sqrt{t_j}|\ffi_j\ket\bra\ffi_k|\otimes \un_C.
\end{align}
\end{prop}
\begin{rem} i) This result applies to the non-Hamiltonian part of the generator of \qrm defined on a simple Hilbert space as well, by considering $\cH_C=\C$, in which case ${\em \tr}_A$ reduces to the scalar valued trace. \\
ii) The operators $A_{jk}$ can be replaced by 
{ 
$\sqrt{t_j}|\ffi_j\ket\bra \psi_k|\otimes \un_C$, 
}where $\{\psi_k\}_k$ is any orthonormal basis of $\cH_A$ without altering the result.
\end{rem}

\subsection{Spectrum of the uncoupled \qrm}
We proceed by analysing the spectrum of the uncoupled \qrm $\cL$ (\ref{triqrm}) in the tri-partite case,
{  
making use of the fact that, by construction, the Hamiltonian part of the decoupled \qrm  commutes with the dissipator as we quickly check:
\begin{align}
 [H_A,\cdot ] \circ (\tau_A\otimes \tr_A(\cdot))(\rho)=[H_A, \tau_A\otimes \tr_A(\rho)]=[H_A,\tau_A]\otimes  \tr_A(\rho)=0,
 \end{align}
 since $\tau_A$ and $H_A$ commute, while
 \begin{align}
(\tau_A\otimes \tr_A(\cdot))\circ [H_A,\cdot ] (\rho)=\tau_A\otimes (\tr_A (H_A\rho)-\tr_A(\rho H_A))=0,
\end{align}
using $\tr_A(\cdot )=\sum_{j}\bra \ffi_j^A|\otimes \un \ \cdot  \ |\ffi_j^A\ket \otimes \un$ with $\{\ffi_j^A\}_{1\leq j\leq n_A}$ an orthonormal basis of eigenvectors of $H_A$.
Now, replacing $H_A$ by $H_B$ (or $H_C$ for that matter) yields
\begin{align}
 &[H_B,\cdot ] \circ (\tau_A\otimes \tr_A(\cdot))(\rho)=\tau_A\otimes [H_B, \tr_A(\rho)], \ \mbox{and} \\ \nonumber
&(\tau_A\otimes \tr_A(\cdot))\circ [H_B,\cdot ] (\rho)=\tau_A\otimes (\tr_A (H_B\rho)-\tr_A(\rho H_B))=\tau_A\otimes [H_B,\tr_A(\rho)],
\end{align}
since $H_B$ commutes with $ \bra \ffi_j^A|\otimes \un$ and $|\ffi_j^A\ket \otimes \un$. Altogether, the dissipator and the Hamiltonian parts of $\cL$ admit a  basis of \aj{common} eigenvectors that we now determine.
\medskip

Let us start }with the dissipator and its spectral properties.
\begin{prop}\label{specdiss} The dissipator, as an operator on $\cB(\cH)$, admits  the following spectral decomposition
\begin{align}\nonumber
&\sigma(\gamma_A(\tau_A\otimes {\em \tr}_A(\cdot )-\un)+\gamma_B({\em \tr}_B(\cdot)\otimes \tau_B-\un))=\{0, -\gamma_A, -\gamma_B, -(\gamma_A+\gamma_B)\}\\ \nonumber
&\gamma_A(\tau_A\otimes {\em \tr}_A(\cdot )-\un)+\gamma_B({\em \tr}_B(\cdot)\otimes \tau_B-\un)=0 Q_0-\gamma_A Q_A-\gamma_B Q_B-(\gamma_A+\gamma_B)Q_{AB},
\end{align}
where the {  spectral} projectors $Q_{\#}$, $\#\in \{0, A, B, AB\}$  are given by
\begin{align}
&Q_0(\rho)=\tau_A\otimes {\em \tr}_{AB}(\rho)\otimes \tau_B, \  \  Q_{AB}(\rho)=\rho-\tr_B(\rho)\otimes \tau_B-\tau_A\otimes\tr_A(\rho)+\tau_A\otimes\tr_{AB}(\rho)\otimes \tau_B, \nonumber \\
&Q_A(\rho)=\big( \tr_B(\rho)-\tau_A\otimes \tr_{AB}(\rho) \big)\otimes\tau_B, \  \ Q_B(\rho)=\tau_A\otimes \big( \tr_A(\rho)- \tr_{AB}(\rho)\otimes \tau_B \big).\nonumber
\end{align}
Moreover, the different spectral subspaces in $\cB(\cH)$ are 
\begin{align}
&\ran Q_0=\spa \{\tau_A\otimes\rho_C\otimes \tau_B\}_{\rho_C\in \cB(\cH_C)}, \hspace{3.7cm} \mbox{s.t.} \   \dim(Q_0)=n_C^2 \nonumber \\ 
&\ran Q_A=\spa \{ \rho_{AC}\otimes \tau_B\, | \, \tr_A(\rho_{AC})=0\}_{\rho_{AC}\in\cB(\cH_A\otimes \cH_C)},  \hspace{0.95cm} \mbox{s.t.} \  \dim(Q_A)=(n_A^2-1)n_C^2\nonumber \\
 &\ran Q_B=\spa \{ \tau_A\otimes \rho_{CB}\, | \, \tr_B(\rho_{CB})=0\}_{\rho_{CB}\in\cB(\cH_C\otimes\cH_B)},  \hspace{1.cm} \mbox{s.t.} \   \dim(Q_B)=(n_B^2-1)n_C^2\nonumber \\
 &\ran Q_{AB}=\spa \{ \tr_A(\rho)=0, \tr_B(\rho)=0 \}_{\rho\in \cB(\cH)},  \hspace{.7cm}  \mbox{s.t.} \   \dim(Q_{AB})= (n_A^2-1)(n_B^2-1)n_C^2.  \nonumber 
\end{align}
\end{prop}
\begin{rem}\label{36} i) In case $\gamma_A=\gamma_B$, there are only three distinct eigenvalues and the corresponding spectral projector is $Q_A+Q_B$.\\
ii) Being spectral projectors, the $Q_{\#}$'s, $\#\in \{0, A, B, AB\}$ satisfy $Q_{0}+Q_{A}+Q_{B}+Q_{AB}=\un$ and $Q_{\#}Q_{\#'}=\delta_{\#,\#'}Q_\#$ \\
iii) The dimensions referred to correspond to complex dimensions for $\cB(\cH)$.\\
iv) The result essentially follows from the observation that  $\tau_A\otimes {\em \tr}_A(\cdot)$ and ${\em \tr}_B(\cdot )\otimes \tau_B$ are commuting projectors.
\end{rem}
\begin{proof}
We start with point iv) of Remark \ref{36}. For any $\rho$ in $\cB(\cH)$, 
\begin{align}
(\tau_A\otimes {\em \tr}_A(\cdot) \circ {\em \tr}_B(\cdot )\otimes \tau_B)(\rho)&=\tau_A\otimes \tr_{AB}(\rho)\otimes \tau_B=({\em \tr}_B(\cdot )\otimes \tau_B\circ \tau_A\otimes {\em \tr}_A(\cdot))(\rho)
\end{align}
while $\tau_A\otimes \tr_A(\cdot)\circ \tau_A\otimes {\em \tr}_A(\cdot)=\tau_A\otimes {\em \tr}_A(\cdot)$, and similarly for $ {\em \tr}_B(\cdot )\otimes \tau_B$. 
Hence the dissipator is a linear combination of two commuting projectors to which we can apply the next Lemma.
\begin{lem}
Let $P, Q \in \cB(\cH)$ such that $P^2=P, Q^2=Q,$ and $[P,Q]=0$. Then, for any $\alpha, \beta\in \C$, the identity
\be
\alpha P+\beta Q=0(\un-P)(\un-Q)+\alpha P(\un-Q)+\beta Q(\un-P)+(\alpha+\beta)PQ,
\ee
provides the spectral decomposition of $\alpha P+\beta Q$, so that $\sigma(\alpha P+\beta Q)=\{0,\alpha, \beta, \alpha+\beta\}$, with respective spectral projectors $(\un-P)(\un-Q), P(\un-Q), Q(\un-P), PQ$, and no eigennilpotent.
\end{lem} The proof of the Lemma is immediate, and in case some eigenvalues coincide, the corresponding spectral projector is simply the sum of the individual projectors.

The identifications $P= \un-\tau_A\otimes {\em \tr}_A(\cdot )$, $Q=\un -{\em \tr}_B(\cdot)\otimes \tau_B$, $\alpha=-\gamma_A$, $\beta=-\gamma_B$ yield the announced spectral decomposition of the dissipator, together with the explicit spectral projectors. A direct  verification then gives the corresponding spectral subspaces. {   \qed}
\end{proof}

\medskip
{  The eigenvectors of the Hamiltonian part of $\cL$ are readily computed.}
For $\#\in\{A,B,C\}$, let $\{\ffi_j^\#\}_{1\leq j\leq n_\#}$ be an orthonormal basis of $\cH_\#$ of eigenvectors of  $H_\#$, with associated eigenvalues $e^\#_j$, $1\leq j\leq n_\#$. The eigenvalues need not to be distinct at that point. We denote by $P^\#_{j,k}\in\cB(\cH_\#)$, $j,k\in\{1,\cdots, n_\#\}$, the operators $P^\#_{j,k}=|\ffi_j^\#\ket\bra\ffi_k^\#|$ that yield a basis of eigenvectors of the Hamiltonian part of (\ref{triqrm}) of  $\cB(\cH))$: 
\begin{align}
-i\big[H_A+H_C  +H_B , &P^A_{j,k}\otimes P^C_{j',k'}\otimes P^B_{j'',k''} \big]\\ \nonumber
&=-i(e^A_j-e^A_k+e^C_{j'}-e^C_{k'}+e^B_{j''}-e^B_{k''})P^A_{j,k}\otimes P^C_{j',k'}\otimes P^B_{j'',k''}.
\end{align}

{   It remains to take into account the role of the trace in the spectral subspaces of the dissipator to get the sought for common basis of eigenvectors of (\ref{dissip}). To do so, we} introduce the $n_\#-1$ dimensional basis of diagonal (w.r.t. to the eigenbasis of $H_\#$) traceless matrices
\begin{align}
\Delta_j^\#=|\ffi_j^\#\ket\bra\ffi_j^\#|-|\ffi_{j+1}^\#\ket\bra\ffi_{j+1}^\#|, \ \ \ j=1, 2, \dots , n_\#-1,
\end{align}
such that $[H_{\#}, \Delta_j^\#]=0$. Together with $\tau_\#$, the $\Delta^\#_j$'s form a basis of diagonal matrices. 
Proposition \ref{specdiss} then provides the full spectral analysis of the uncoupled \qrm.
\begin{prop}\label{specunc} The vectors listed below form a basis of $\cB(\cH)$ consisting in eigenvectors associated with the mentioned eigenvalue of   
the uncoupled \qrm \\ $\cL(\cdot)=-i\big[H_A+H_C  +H_B , \cdot  \big]+\cD(\cdot)$ defined on $\cH=\cH_A\otimes\cH_C\otimes\cH_B$ by (\ref{triqrm}):
\begin{align}
& \tau_A\otimes P^C_{j',k'}\otimes\tau_B\,\leftrightarrow \,-i(e^C_{j'}-e^C_{k'}), \nonumber \\
&\hspace{ 12.2cm}    1\leq j', k' \leq n_C\nonumber \\
& \Delta_j^A\otimes P^C_{j',k'}\otimes\tau_B \,\leftrightarrow \, -\gamma_A-i(e^C_{j'}-e^C_{k'}), \nonumber \\
&\hspace{ 9.3cm}  1\leq j\leq n_A-1, \ 1\leq j', k' \leq n_C\nonumber \\
& P^A_{j,k}\otimes P^C_{j',k'}\otimes\tau_B\,\leftrightarrow \, -\gamma_A-i(e^A_j-e^A_k+e^C_{j'}-e^C_{k'}), \nonumber \\
&\hspace{ 9.3cm}   1\leq j \neq k \leq n_A, \ 1\leq j', k' \leq n_C\nonumber \\
& \tau_A\otimes P^C_{j',k'}\otimes \Delta_{j''}^B\,\leftrightarrow \, -\gamma_B-i(e^C_{j'}-e^C_{k'}), \nonumber \\
&\hspace{ 9.15cm}   1\leq j''\leq n_B-1, \ 1\leq j', k' \leq n_C\nonumber \\
& \tau_A\otimes P^C_{j',k'}\otimes P^B_{j'',k''}\,\leftrightarrow \, -\gamma_B-i(e^C_{j'}-e^C_{k'}+e^B_{j''}-e^B_{k''}), \nonumber \\
&\hspace{ 8.9cm}  1\leq j'' \neq k'' \leq n_B, \ 1\leq j', k' \leq n_C\nonumber \\
& \Delta^A_{j}\otimes P^C_{j',k'}\otimes \Delta^B_{j''}\,\leftrightarrow \, -(\gamma_A+\gamma_B)-i(e^C_{j'}-e^C_{k'}), \nonumber \\
&\hspace{ 6.2cm} 1\leq j \leq  n_A-1, \  1\leq j'' \leq n_B-1, \ 1\leq j', k' \leq n_C\nonumber \\
& \Delta^A_{j}\otimes P^C_{j',k'}\otimes P^B_{j'',k''}\,\leftrightarrow \, -(\gamma_A+\gamma_B)-i(e^C_{j'}-e^C_{k'}+e^B_{j''}-e^B_{k''}), \nonumber \\
&\hspace{ 6cm} 1\leq j \leq n_A-1, \  1\leq j'' \neq k'' \leq n_B, \ 1\leq j', k' \leq n_C\nonumber \\
& {  P^A_{j,k}\otimes P^C_{j',k'}\otimes \Delta^B_{j''}\,\leftrightarrow \, -(\gamma_A+\gamma_B)-i(e^A_j-e^A_k+e^C_{j'}-e^C_{k'}), } \nonumber \\
&\hspace{6.2cm} 1\leq j \neq k \leq n_A, \  1\leq j'' \neq  n_B-1, \ 1\leq j', k' \leq n_C\nonumber \\
& P^A_{j,k}\otimes P^C_{j',k'}\otimes P^B_{j'',k''}\,\leftrightarrow \, -(\gamma_A+\gamma_B)-i(e^A_j-e^A_k+e^C_{j'}-e^C_{k'}+e^B_{j''}-e^B_{k''}), \nonumber \\
&\hspace{ 6cm} 1\leq j \neq k \leq n_A, \  1\leq j'' \neq k'' \leq n_B, \ 1\leq j', k' \leq n_C\nonumber 
\end{align}
\end{prop}
\begin{rem} 0) The Hamiltonians $\cH_\#\in\cB(\cH_\#)$ are arbitrary at that point.\\
i) The uncoupled reset model Lindbladian $\cL$ is thus diagonalisable, with eigenvalues located on the (generically) four vertical lines $\Re z=0$, $\Re z=-\gamma_A$, $\Re z=-\gamma_B$, $\Re z=-(\gamma_A+\gamma_B)$ in the complex plane, symmetrically with respect to the real axis.\\
ii) In particular, the kernel of $\cL$ is degenerate, since $\,\dim\ker \cL(\cdot)\geq n_C$.\\
\aj{iii) It is straightforward to generalise this result to the case where the dissipator admits a reset part acting on $\cH_C$ as well, and to the case of a $p$-partite non interacting system, with $p\in\N$ arbitrary. }
\end{rem}

The spectral projectors of $\cL$ can be constructed explicitly, making use of the next Lemma:

\begin{lem}\label{specpro}
Consider a Hilbert space $\cH$ and $\tau\in \cB(\cH)$ a density matrix. Let  $\{\ffi_j\}_{1\leq j\leq n}$  be an orthonormal basis of eigenvectors of $\tau$ for $\cH$.  Consider the basis of  $\cB(\cH)$ given by
 \be\label{taubasis}
 P_{jk}=|\ffi_j\ket\bra\ffi_k|, \ 1\leq j\neq k\leq n, \ \Delta_j=|\ffi_j\ket\bra\ffi_j|-|\ffi_{j+1}\ket\bra\ffi_{j+1}|, \  1\leq j\leq n-1, \  \mbox{and } \ \tau.
 \ee
Set $\sigma_j=\sum_{k=1}^{j}|\ffi_k\ket\bra\ffi_k|$, $1\leq j\leq n$. Then the operators on $\cB(\cH)$ defined by
\begin{align}\label{projL}
&Q_{jk}(\cdot)=P_{jk}\tr(P_{jk}^*\, \cdot\, ), \ 1\leq j\neq k \leq n, \nonumber \\
&Q_j(\cdot)=\Delta_j \tr(\sigma_j (\,\cdot -\tau\tr(\cdot))), \ 1\leq j\leq n-1, \ \mbox{and} \ \ Q_0(\cdot)=\tau \tr (\un \, \cdot)
\end{align}
yield a complete set of rank one projectors onto the span of the corresponding basis vectors of (\ref{taubasis}) so that the composition of any two of them equals zero.
\end{lem}
\begin{rem} 
The spectral projectors of $\cL$ corresponding to Proposition \ref{specunc} are then given by the appropriate tensor products of projectors (\ref{projL}).
\end{rem}

The solution to $\dot \rho=\cL(\rho)$, $\rho(0)=\rho_0$ follows immediately by expanding $\rho_0$ along those eigenvectors. In particular, one gets for this uncoupled \qrm model
\be
\rho(t)=\tau_A\otimes (e^{-i[H_C,\cdot]t}\tr_{AB}(\rho_0))\otimes \tau_B+\ode(e^{-t\min\{\gamma_A,\gamma_B\}}), \ \ t\geq 0,
\ee
where $e^{-i[H_C,\cdot]t}\tr_{AB}(\rho_0)$ satisfies the Hamiltonian evolution equation $\dot\rho_C=-i[H_C,\rho_C]$, $\rho_C(0)=\tr_{AB}(\rho_0)$ on $\cH_C$, as expected in this uncoupled context.

\section{The weakly-interacting tripartite \qrm}
\label{sec:3}

We consider now the coupled \qrm defined by the Lindblad generator on $\cB(\cH)$, with $\cH=\cH_A\otimes\cH_C\otimes\cH_B$,
\be\label{couqrm}
\cL_g(\rho)=\cL(\rho)-ig[H,\rho]\equiv \cL_0(\rho)+g\cL_1(\rho)
\ee
where $H=H^* \in \cB(\cH)$ is a Hamiltonian that effectively couples the different Hilbert spaces $\cH_\#$, while $g\in \R$ is a coupling constant. We focus on the determination of the kernel of $\cL_g$, 
as $g\ra 0$, which describes the asymptotic state of the system driven by $\cL_g$, under generic hypotheses.
Then we turn to the consequences for the dynamics generated by $\cL_g$.
 \aj{By generic hypotheses, we mean that all assumptions we make along the way 
ensure the coupling is effective enough to lift all degeneracies, so that all accidental degeneracies are eliminated order by order in $g$.}

\subsection{Leading order analytic perturbation theory}

When $g=0$, Proposition \ref{specunc} shows that 
\be 
\ker \cL_0\supset \spa \big\{\tau_A\otimes |\ffi_j^C\ket\bra \ffi_j^C|\otimes \tau_B\big\}_{1\leq j\leq n_C},
\ee 
whatever the properties of the Hamiltonian $H_C$. \aj{We shall consider below both cases $H_C=0$ and $H_C\neq 0$, which give rise to different results.} In case the Hamiltonian $H_C$ is trivial, 
\be
H_C=0 \ \Rightarrow \ \ker \cL_0= \spa \big\{\tau_A\otimes \rho_C\otimes \tau_B\big\}_{\rho_C\in\cB(\cH_C)}
\ee has dimension $n_C^2$, and the corresponding spectral projector coincides with $Q_0$, the spectral projector on $\ker  \cD$, see Proposition \ref{specdiss}. 
In order to avoid accidental degeneracies when $H_C\neq 0$, we will assume $H_C$ satisfies the spectral hypothesis\\

\noindent
{\bf Spec($H_C$)}: 
\\The spectrum of $H_C\in\cB(\cH_C)$ is simple and the Bohr frequencies $\{e^C_j-e_k^C\}_{j\neq k}$ are distinct. \\

Under this assumption, we have 
\be
\ker  \cL_0=\spa \big\{\tau_A\otimes \rho_C \otimes \tau_B, \ \mbox{s.t.} \ [\rho_C,H_C]=0\big\},
\ee which is of dimension $n_C$. The corresponding spectral projector 
acts as follows 
\be\label{q0hc}
Q_0(\rho)=\tau_A\otimes \diag_C(\tr_{AB}(\rho))\otimes \tau_B,
\ee
where the projector $\diag_C : \cB(\cH_C)\ra \cB(\cH_C)$ defined by
\be
\diag_C (\cdot )=\sum_{j=1}^{n_C} | \ffi_j^C\ket\bra \ffi_j^C| \ \cdot \   | \ffi_j^C\ket\bra \ffi_j^C|
\ee
extracts the diagonal part of $\rho_C$ within the normalised eigenbasis of $H_C$. Observe that  $\offdiag_C : \cB(\cH_C)\ra \cB(\cH_C)$, extracting the offdiagonal part of $\rho_C$ within the same basis, yields the complementary projector 
\be
\offdiag_C=\un -\diag_C.
\ee
We also note, for later reference, that $Q_0$ on $\cB(\cH)$ is trace preserving, so that $\ran (\un-Q_0)\subset \{\rho \, | \, \tr \rho =0\}$.

\medskip

Analytic perturbation theory, see {e.g.}\,Chapter II \S 2 \cite{K}, allows us to compute the splitting of the degenerate eigenvalue zero of $\cL_0$ by the perturbation $g\cL_1$. Recall here that $\cL_g$ being a Lindblad operator (Proposition \ref{lind}), the following structural constraints hold:
\be\label{speclind}
0\in \sigma(\cL_g)=\overline{\sigma(\cL_g)}\subset\{z\in \C \ | \Re z\leq 0\}, \ \ \forall \ g\in \R.
\ee 
Moreover, the eigenvalue $0$ is semisimple, that is there is no eigennilpotent (Jordan block) corresponding to that eigenvalue in the spectral decomposition of $\cL_g$. The same is actually true for all eigenvalues sitting on the imaginary axis.
\medskip

Let $\{\lambda_j(g)\}_{1\leq j\leq m}$ be the set of eigenvalues of $\cL_g$ that stem from the eigenvalue $0$ of $\cL_0$, with $m=n_C^2$ if $H_C=0$ or $m=n_C$ if $H_C\neq 0$. They form the so-called $\lambda-$group for $\lambda=0$, and for $g\in \C\setminus\{0\}$ with $|g|$ is small enough, $\{\lambda_j(g)\}_{1\leq j\leq m}$ are analytic functions of a (fractional) power of $g$ that tend to zero as $g\ra 0$.
These eigenvalues may be permanently degenerate. For the structural reasons recalled above, one of these eigenvalues, denoted by $\lambda_0(g)$, is identically equal to zero, $\lambda_0(g)\equiv 0$, $\forall g\in\C\setminus{0}$, and in case $\lambda_0(g)$ is degenerate, it is semisimple.

We show that under generic hypotheses, $\lambda_0(g)\equiv 0$ is a simple eigenvalue, see Theorem \ref{goldenrule}, and we determine the corresponding eigenvector $\rho_0(g)$, normalized to be a state, {\it i.e.} $\rho_0(g)\geq 0$ and $\tr \rho_0(g)= 1$.

\medskip

Let us denote by $Q_0(g)$ the analytic spectral projector of $\cL_g$ corresponding to the set of eigenvalues in the  $0$-group . It writes 
\be\label{0gpespero}
Q_0(g)=\frac{-1}{2i\pi}\int_{\Gamma_0}(\cL_g-z)^{-1}dz=Q_0+gQ_1+g^2Q_2+\ode(g^3), 
\ee 
for $|g|$ is small, where  $\Gamma_0$ is a circle of small radius centered at the origin. Also, since $0$ is a semisimple eigenvalue of $\cL_0$, 
\be\label{q1}
Q_1=- Q_0\cL_1 S_0-S_0\cL_1 Q_0=Q_0Q_1(\un-Q_0)+(\un-Q_0)Q_1Q_0,
\ee
 where $S_0$ is the reduced resolvent of $\cL_0$ at $0$, satisfying $S_0Q_0=Q_0S_0=0$ and $S_0\cL_0=\cL_0S_0=\un-Q_0$. In other words,
 $S_0=\cL_0^{-1}(\un -Q_0)$, that we shall sometimes abusively write $S_0=\cL_0^{-1}$, with the understanding that it acts on $(\un -Q_0)\cB(\cH)$.
 The analytic reduced operator in the corresponding subspace which describes the splitting reads
\begin{align}
 Q_0(g)\cL_gQ_0(g) 
&=(Q_0+gQ_1+g^2Q_2+\ode(g^3))(\cL_0+g\cL_1)(Q_0+gQ_1+g^2Q_2+\ode(g^3)) \nonumber \\
&= gQ_0 \cL_1 Q_0 + g^2 (Q_1 \cL_0 Q_1+Q_1\cL_1Q_0+ Q_0 \cL_1 Q_1)+ \ode(g^3),
\end{align}
where we used $\cL_0Q_0=Q_0\cL_0=0$.

\medskip

\begin{lem}\label{redop} Under assumption {\bf Spec($H_C$)} when $H_C\neq 0$, we have  
\begin{align}
Q_0 \cL_1 Q_0(\rho)=\left\{\begin{matrix} 0 & \mbox{if } \ H_C\neq 0 \cr
-i\tau_A\otimes [\overline{H}^{\, \tau},  \tr_{AB}(\rho)]\otimes \tau_B &  \mbox{if } \ H_C= 0 \end{matrix} \right.
\end{align}
where 
\begin{align}\label{hbar}
\overline{H}^{\, \tau} :&=  \tr_{AB}(\tau_A^{1/2}\otimes \un_C \otimes\tau_B^{1/2}\, H\, \tau_A^{1/2}\otimes \un_C \otimes\tau_B^{1/2})\\ \nonumber
&=\tr_{AB}(H\, \tau_A\otimes \un_C \otimes\tau_B)=\tr_{AB}(\tau_A\otimes \un_C \otimes\tau_B \,H)\in\cB(\cH_C).
\end{align}
\end{lem}
Explicitly, with $\tau_\#=\sum_{1\leq j\leq n_\#}t_j^\# | \ffi_j^\#\ket \bra \ffi_j^\#|$,
\be
\overline{H}^{\, \tau}=\sum_{1\leq j\leq n_A \atop 1\leq k\leq n_B}t_j^At_k^B (\bra \ffi_j^A | \otimes \un_C\otimes \bra \ffi_k^B|) \ H \ (| \ffi_j^A \ket \otimes \un_C \otimes  |\ffi_k^B\ket) .
\ee
As a consequence, when $H_C\neq 0$ the splitting is generically described by the order $g^2$ correction, while in case $H_C=0$, the non-zero first order correction imposes that the elements of the kernel of 
$Q_0(g)$ commute with $\overline{H}^{\, \tau}$ which, generically, decreases the degeneracy from $n_C^2$ to $n_C$. In both cases, the eigenvalue zero of $Q_0 \cL_1 Q_0$ is 
semisimple.\\

\begin{proof}
We first compute for any $\rho_C\in\cB(\cH_C)$, using (\ref{hbar}), 
\begin{align}
\tr_{AB}([H,\tau_A\otimes \rho_C\otimes\tau_B])=[\overline{H}^{\, \tau}, \rho_C].
\end{align}
One gets the explicit expression for $\overline{H}^{\, \tau}$ by expressing the partial trace within the eigenbases of $\tau_\#$.
Therefore
\be
Q_0 \cL_1 Q_0(\rho)=-i\tau_A\otimes (\tr_{AB}([H,\tau_A\otimes \tr_{AB}(\rho)\otimes\tau_B]))\otimes \tau_B= -i\tau_A\otimes [\overline{H}^{\, \tau},  \tr_{AB}(\rho)]\otimes \tau_B.
\ee
The fact that  $H_C\neq 0$ implies $Q_0 \cL_1 Q_0=0$ then follows from 
\be
Q_0 \cL_1 Q_0(\rho)=-i\tau_A\otimes \diag_C(\tr_{AB}([H,\tau_A\otimes \diag_C(\tr_{AB}(\rho))\otimes\tau_B]))\otimes \tau_B,
\ee
and the identity
\be
\diag_C (\tr_{AB}([H,\tau_A\otimes \diag_C(\rho_C)\otimes\tau_B]))=\diag_C [\overline{H}^{\, \tau}, \diag_C (\rho_C)]=0.
\ee
\end{proof} \hfill $\Box$

Let us investigate the next order correction in order to analyse the splitting from the eigenvalue zero. Following \cite{K} we consider the analytic matrix
\begin{align}\label{tildel}
\widetilde \cL_g&=\frac{1}{g}Q_0(g) \cL_g Q_0(g) \nonumber \\ \nonumber
&= Q_0 \cL_1 Q_0 + g (Q_1 \cL_0 Q_1+Q_1\cL_1Q_0+ Q_0 \cL_1 Q_1)+ \ode(g^2) \\
&\equiv \widetilde \cL_0+g \widetilde \cL_1 + \ode(g^2),
\end{align}
where we observe with (\ref{q1}) that
\begin{align}
 \widetilde \cL_1=- Q_0 \cL_1S_0\cL_1 Q_0 - S_0\cL_1 Q_0 \cL_1 Q_0-Q_0\cL_1 Q_0 \cL_1 S_0.
\end{align}
Let $\widetilde Q_0$ be the eigenprojector onto $\ker  \widetilde \cL_0$. Then  the spectrum of $\widetilde Q_0  \widetilde \cL_1\widetilde Q_0$ 
describes the splitting to order $g^2$, see \cite{K}, Thm 5.11:  for $\tilde \lambda_j^{(1)}\in \sigma (\widetilde Q_0  \widetilde \cL_1\widetilde Q_0)$ 
of multiplicity $m_j^{(1)}$,  there exist exactly 
$m_j^{(1)}$ eigenvalue of 
$\cL_g$ of the form 
\be
\lambda_j(g)=g^2\tilde \lambda_j^{(1)}+ \ode(g^3).
\ee
Notice that $\widetilde Q_0  \widetilde \cL_1\widetilde Q_0$ is viewed as an operator on $Q_0\cB(\cH)$ here. 

We observe that 
$\widetilde Q_0=\widetilde Q_0 Q_0  =Q_0  \widetilde Q_0$, hence
\begin{align}
\widetilde Q_0  \widetilde \cL_1\widetilde Q_0=-\widetilde Q_0 ( Q_0\cL_1S_0\cL_1 Q_0  )\widetilde Q_0=-\widetilde Q_0 \cL_1\cL_0^{-1}\cL_1 \widetilde Q_0,
\end{align}
since $\cL_1 \widetilde Q_0=(\un -Q_0)\cL_1 \widetilde Q_0$.

In order to proceed, we shall also assume in the sequel that  the operator $\overline{H}^{\, \tau}$ appearing in 
Lemma \ref{redop} has generic spectral properties.\\

\noindent 
{\bf Spec($\overline{H}^{\, \tau}$)}:\\
\noindent 
The spectrum of $\overline{H}^{\, \tau}\in\cB(\cH_C)$ is simple and the corresponding Bohr frequencies are distinct. 
We denote the normalised eigenvectors and eigenvalues of $\overline{H}^{\, \tau}$ by $\ffi_j^\tau$ and $e_j^\tau$, $1\leq j\leq n_C$.\\

Under {\bf Spec($\overline{H}^{\, \tau}$)}, we get from (\ref{q0hc}) and Lemma \ref{redop}
\begin{align}
\widetilde Q_0(\rho)=\left\{\begin{matrix}
\tau_A\otimes \diag_C \tr_{AB}(\rho) \otimes \tau_B & \ \mbox{if} \, H_C\neq 0\phantom{,} \\
\tau_A\otimes \diag_\tau \tr_{AB}(\rho) \otimes \tau_B & \  \mbox{if} \, H_C= 0, 
\end{matrix}\right.\end{align}
where $\diag_\tau$ is the projector that extracts the diagonal part of the matrices expressed in the orthonormal eigenbasis $\{\ffi_j^\tau\}$.
Therefore
\be
\widetilde Q_0  \widetilde \cL_1\widetilde Q_0=\tau_A\otimes \diag\, \tr_{AB}\Big( \big[H, \cL_0^{-1}([H, \tau_A\otimes \diag\, \tr_{AB}(\cdot ) \otimes \tau_B])\big]\Big) \otimes \tau_B,
\ee
where $\diag$ stands here for $\diag_C$ (resp. $\diag_\tau$) if  $H_C\neq 0$ (resp. $H_C=0$). Equivalently, $\widetilde Q_0  \widetilde \cL_1\widetilde Q_0$ is fully characterised by 
the following linear map. Set
\be\label{phi}
\Phi(\cdot):= {\em \tr}_{AB}\big(\big[H, \cL_0^{-1}([H, \tau_A\otimes \diag(\, \cdot \,) \otimes \tau_B])\big]\big) : \cB(\cH_C)\ra \cB(\cH_C)\cap \{\rho_C\, | \, \tr \rho_C=~0\}.
\ee
Note that $\Phi$ is well defined and takes the form  $\Phi(\rho)=\tr_{AB} ([H, M(\rho)])$, for $M(\rho)\in \cB(\cH)$,
hence 
$\tr\,  \Phi(\rho)=\tr ([H,M(\rho)])=0$, for any $\rho$. Then, the restriction of $\Phi$ to $ \diag \, \cB(\cH_C)$, which has dimension $n_C$, satisfies 
\begin{align}\label{deffid}
\Phi_D:=\diag\,  \Phi \,  |_{\diag \, \cB(\cH_C)} \ \ \mbox{and } \ \ \widetilde Q_0  \widetilde \cL_1\widetilde Q_0(\cdot)=\tau_A\otimes \Phi_D(\cdot)\otimes \tau_B\circ \tr_{AB}(\cdot).
\end{align}
We shall abuse notations in the sequel and simply write 
\be
\widetilde Q_0  \widetilde \cL_1\widetilde Q_0(\cdot)=\tau_A\otimes \Phi_D(\cdot)\otimes \tau_B,
\ee
identifying operators defined on $\widetilde Q_0\cB(\cH)$ and $\diag \, \cB(\cH_C)$. Hence 
\be
 \sigma(\widetilde Q_0  \widetilde \cL_1\widetilde Q_0)=\sigma(\Phi_D).
\ee
Note that $\dim \ker\,  \Phi_D\geq 1$, since $ \ran \,  \Phi_D\subset  \diag \, \cB(\cH_C)\cap \{\rho_C\, | \, \tr \rho_C=~0\}$, a subspace of dimension $n_C-1$, in keeping
with the fact that $\ker  \cL_g$ is never trivial. Hence, for the zero eigenvalue of $\cL_g$ to be non-degenerate at second order perturbation in $g$, we assume the coupling satisfies the assumption. \\

\noindent
{\bf Coup}:\\
The linear map 
\be\label{defphi}
\Phi_D (\cdot)=\diag\,  {\em \tr}_{AB}\big(\big[H, \cL_0^{-1}([H, \tau_A\otimes \diag\, (\, \cdot \,) \otimes \tau_B])\big]\big)  \ \ \mbox{defined on} \  \diag \, \cB(\cH_C),
\ee
where $\diag$ stands here $\diag_C$ (resp. $\diag_\tau$) if  $H_C\neq 0$ (resp. $H_C=0$), is such that $\dim \ker\, \Phi_D=1$.
\begin{rem} Assumption {\em \bf Coup} is equivalent to the statement \vspace{-.2cm } \\ {\em

\noindent
$\Phi_D^{-1}$ exists on the $n_c-1$ dimension  subspace $ \diag \, \cB(\cH_C)\cap \{\rho_C\, | \, \tr \rho_C=~0\}=~\ran \Phi_D$.} \vspace{-.3cm } \\
 
 Indeed, both statements entail $\dim \ker \Phi_D=1$, and
the corresponding spectral projector onto $\ker  \Phi_D$, say $\Pi_0$, is such that $\ran \Phi_D=(\un - \Pi_0)\ran \,\Phi_D$, and $\ker  \Phi_D \cap \ran \Phi_D=\{ 0\}$. 
\end{rem}

As a consequence,
\begin{thm}\label{goldenrule} Consider the coupled \qrm $\cL_g(\rho)$ defined on $\cB(\cH_A\otimes\cH_C\otimes\cH_B)$ by 
\begin{align}\label{lgthm}
\cL_g(\rho)=&-i\big[H_A+H_C  +H_B +gH, \rho\big]+\gamma_A(\tau_A\otimes \tr_A(\rho)-\rho)+\gamma_B(\tr_B(\rho)\otimes \tau_B-\rho)
\end{align}
and assume {\bf Spec($H_C$)} if $H_C\neq 0$ or  {\bf Spec($\overline{H}^{\, \tau}$)}  if  $H_C=0$. Then for $g\in\C\setminus {\{0\}}$, $|g|$ small enough,  $\dim \ker\, \cL_g=1$ if {\bf Coup} holds. 
\end{thm}

\begin{rem}\label{nzspec}
Under assumption  {\bf Spec($\overline{H}^{\, \tau}$)}, the non-zero eigenvalues of $\widetilde \cL_0$ are all simple, of the form $ \lambda_{jk}=-i(e_j^\tau-e_k^\tau)$ with associated eigenvector 
$\tau_A\otimes|\ffi_j^\tau\ket\bra \ffi_k^\tau |\otimes \tau_B$, $j\neq k$, and corresponding eigenprojector 
\be
\widetilde Q_{\lambda_{jk}}(\rho)=\tau_A\otimes|\ffi_j^\tau\ket\bra \ffi_k^\tau |\otimes \tau_B \ \tr (\tau_A\otimes|\ffi_k^\tau\ket\bra \ffi_j^\tau |\otimes \tau_B \, \rho).
\ee
The next order correction, given by the eigenvalue of the operator $-\widetilde Q_{\lambda_{jk}} \cL_1\cL_0^{-1}\cL_1\widetilde Q_{\lambda_{jk}}$, reads
\begin{align}\label{simplenz}
\tilde \lambda_{jk}^{(1)}=\tr \Big\{(\un_A\otimes|\ffi_k^\tau\ket\bra \ffi_j^\tau |\otimes \un_B)\big[H,\cL_0^{-1}([H,\tau_A\otimes|\ffi_j^\tau\ket\bra \ffi_k^\tau |\otimes \tau_B])\big]\Big\}.
\end{align}
\end{rem}

\subsection{Dynamics}

We push here the spectral analysis a bit further in order to get sufficient information to analyse the behaviour of the dynamics of the coupled \qrm $\cL_g(\cdot)$,
as $g\ra 0$. We first discuss the richer case $H_C=0$ and then describe the modifications required for the case $H_C\neq 0$.

\medskip

Let $Q_0(g)$ be the spectral projector of $\cL_g$ given by (\ref{0gpespero}), and $Q_0^\flat(g)=\un - Q_0(g)$. We have accordingly
\be
e^{t\cL_g}=e^{t\cL_g^{0}}Q_0(g)+e^{t\cL_g^{\flat}}Q_0^\flat(g), 
\ee
where $\cL_g^0=\cL_g|_{\ran Q_0(g)}$, and  $\cL_g^{\flat}=\cL_g|_{\ran Q_0(g)}$. Since the spectrum of $\cL_g^{\flat}$ is a positive distance away from the imaginary axis, uniformly in $g$ small enough, functional calculus yields the existence of $\Gamma>0$, independent of $g$, such that
\be\label{easy}
e^{t\cL_g}=e^{t\cL_g^{0}}Q_0(g)+\ode (e^{-t\Gamma}), 
\ee
where $\ode $ is uniform in $g$, since $Q_0(g)$ is analytic in $g$. 
Now, by (\ref{tildel}) 
\be\label{renormsplit}
\cL_g^0=g\widetilde \cL_g=g(\widetilde \cL_0+g\widetilde \cL_1+\ode (g^2)),
\ee
where, for $H_C=0$ under assumption  {\bf Spec($\overline{H}^{\, \tau}$)},
\be\label{declt0}
\widetilde \cL_0=0 \,\widetilde Q_0 + \sum_{j\neq k} \lambda_{jk}\widetilde Q_{\lambda_{jk}}, \ \mbox{where } \ \lambda_{jk}=-i(e^\tau_j-e_k^\tau),
\ee
with simple non zero eigenvalues, see Remark \ref{nzspec}. 
In case $H_C\neq 0$ under hypothesis  {\bf Spec($H_C$)},  $\widetilde \cL_0=0$ by Lemma \ref{redop}, so that (\ref{declt0}) holds with $\widetilde Q_0= Q_0$ and $\widetilde Q_{\lambda_{jk}}=0$.

Since $\cL_g^0=\ode(g)$ (and even $\cL_g^0=\ode(g^2)$ in case $H_C\neq 0$), the long time behaviour of $e^{t\cL_g}$ is controlled by the first term in (\ref{easy}) when $g$ is small. This requires addressing the behaviour of the non self-adjoint spectral projectors associated to eigenvalues of $\cL_g$ that vanish as $g$ goes to zero. 
\begin{prop} \label{spec2}
Assuming $H_C=0$, {\bf Spec($\overline{H}^{\, \tau}$)} and  {\bf Coup}, there exists $g_0>0$ such that for all $|g|<g_0$, $\cL_g$ admits analytic spectral projector $\widetilde Q_0(g) $ and $\widetilde Q_{\lambda_{jk}}(g)$
 and analytic simple eigenvalues $\lambda_{jk}(g)$ such that
\be\label{firstdec}
\cL_g^0= g\widetilde Q_0(g)\widetilde \cL_g\widetilde Q_0(g) +\sum_{j\neq k}\lambda_{jk}(g)\widetilde Q_{\lambda_{jk}}(g).
\ee
Here $\widetilde Q_0(g) =\widetilde Q_0+ \ode(g)$, $\widetilde Q_{\lambda_{jk}}(g)=\widetilde Q_{\lambda_{jk}}+\ode(g)$ and $\lambda_{jk}(g)=-ig(e^\tau_j-e_k^\tau)+g^2\tilde \lambda_{jk}^{(1)}+\ode (g^3)$, see (\ref{simplenz}).

Assuming $H_C\neq 0$, {\bf Spec($H_C$)} and {\bf Coup}, the same statement holds with $\widetilde Q_{0}(g)=Q_0+\ode(g)$ and $\widetilde Q_{\lambda_{jk}}(g)\equiv 0$, $\lambda_{jk}(g)\equiv 0$   in (\ref{firstdec}).

Moreover, assuming  {\bf Coup} and  {\bf Spec($\overline{H}^{\, \tau}$)}, (respectively  {\bf Spec($H_C$)}), if  $H_C=0$, (respectively  $H_C\neq 0$), we have $\dim \ker \widetilde Q_0(g)\widetilde \cL_g\widetilde Q_0(g)\equiv 1$ and the corresponding  spectral projector $\widetilde Q_0^S(g)$ is analytic for $|g|<g_0$, and satisfies
\be \label{q0s}
\widetilde Q_0^S(g)\cL_g=\cL_g \widetilde Q_0^S(g)=0.
\ee 
Here 
\be
\widetilde Q_0(g)\widetilde \cL_g\widetilde Q_0(g)=g\widetilde Q_0\widetilde \cL_1\widetilde Q_0+\ode(g^2)= g \tau_A\otimes \Phi_D \otimes \tau_B+\ode(g^2)
\ee
 and $\widetilde Q_0^S(g)=\widetilde Q_0^S+\ode(g)$ where $\widetilde Q_0^S$ is the projector on $\ker \widetilde Q_0\widetilde \cL_1\widetilde Q_0$.
 \end{prop}
\begin{rem}
The spectral constraints on Lindblad operators imply,
\be\label{constspec}
\Re \, \sigma(\widetilde Q_0\widetilde \cL_1\widetilde Q_0)\setminus{\{0\}} \leq 0, \ \mbox{ and } \ \Re \tilde \lambda_{jk}^{(1)} \leq 0.
\ee
We give conditions ensuring $\Re \tilde \lambda_{jk}^{(1)}<0$ in  case the model has no leading order Hamiltonian drive, $\cL_0=\cD$, that we analyse in more details in Section \ref{secnodrive}. 
\end{rem}

\begin{proof}We consider $H_C=0$ only, the other case being similar.
Thanks to (\ref{renormsplit}) and (\ref{declt0}), perturbation theory applies to $\widetilde \cL_g$ and yields the analytic projectors 
$\widetilde Q_0(g) $ and $\widetilde Q_{\lambda_{jk}}(g)$ converging to $\widetilde Q_0$ and $\widetilde Q_{\lambda_{jk}}$ respectively, and the analytic simple eigenvalues 
$\lambda_{jk}(g)$, such that (\ref{firstdec}) holds.
Expanding the first term using $\widetilde Q_0\widetilde \cL_0= \widetilde \cL_0\widetilde Q_0=0$, one gets thanks to (\ref{deffid})
 \be
 \widetilde Q_0(g)\widetilde \cL_g\widetilde Q_0(g)= g \widetilde Q_0\widetilde \cL_1\widetilde Q_0+\ode(g^2)=g\, \tau_A\otimes \Phi_D \otimes \tau_B+\ode(g^2).
 \ee
Assumption {\bf Coup} implies that $\widetilde Q_0\widetilde \cL_1\widetilde Q_0$ has one dimensional kernel, with associated spectral projector we write $\widetilde Q_0^S$. Hence, perturbation theory again ensures the existence of an analytic one dimensional  spectral projector $\widetilde Q_0^S(g)$ of $\widetilde Q_0(g)\widetilde \cL_g\widetilde Q_0(g)$ corresponding to the simple zero eigenvalue of  $\widetilde Q_0\widetilde \cL_1\widetilde Q_0$ at $g=0$. Necessarily, $\widetilde Q_0^S(g)$ coincides with the spectral projector onto the nontrivial  kernel of $\cL_g$ for all $g$ small enough, which proves (\ref{q0s}). 
 $\hfill \Box$
\end{proof}
\medskip

Let us turn to the dynamical implications.
\begin{cor}\label{dynimp} Under the hypotheses for $H_C=0$ above, the following holds \aj{for all $t\geq 0$ and} $g$ real small enough:
 \be
e^{t\cL_g}=  e^{tg^2(\widetilde Q_0\widetilde \cL_1\widetilde Q_0+\ode(g))}\widetilde Q_0(g)+ \sum_{j\neq k} e^{t\lambda_{jk}(g)}\widetilde Q_{\lambda_{jk}}(g)+\ode(e^{-t\Gamma}).
\ee
Further assuming $\max_{j \neq k} \{\Re \tilde \lambda_{jk}^{(1)}\} <0$, there exists $\delta >0$ such that for all $t\geq 0$,
\begin{align}\label{id1}
e^{t\cL_g}&=\widetilde Q_0^S(g)+\ode (e^{-\delta g^2 t}),
\end{align}
where the constant in the $\ode $ is uniform in $t\geq 0$ and $g$ small. 

Setting $\eta = \min_{j \neq k} \{|\Re \tilde \lambda_{jk}^{(1)}|\}>0$, we have
\begin{align}\label{appdyn}
e^{t\cL_g}&=e^{tg^2\widetilde Q_0\widetilde \cL_1\widetilde Q_0}\widetilde Q_0+ \ode(e^{-tg^2\eta}) + \ode (g)+\ode (g^3 t)\nonumber \\
&=\tau_A\otimes e^{tg^2\Phi_D}\diag_\tau \tr_{AB} \otimes {\tau_B} + \ode(e^{-tg^2\eta}) + \ode (g)+\ode (g^3 t)
\end{align}
where the constants in all $\ode$ are uniform in $t\geq 0$, $g$ small.

Under the hypotheses for $H_C\neq 0$ above, \aj{for all $t\geq 0$ and} $g$ real small enough, 
\be
e^{t\cL_g}=  e^{tg^2(Q_0\widetilde \cL_1 Q_0+\ode(g))}\widetilde Q_0(g)+\ode(e^{-t\Gamma}),
\ee
and there exists $\delta >0$ such that for all $t\geq 0$,
\begin{align}\label{id2}
e^{t\cL_g}&=\widetilde Q_0^S(g)+\ode (e^{-\delta g^2 t}),
\end{align}
where the constant in the $\ode $ is uniform in $t\geq 0$ and $g$ small. Moreover,
\begin{align}\label{appdynhc}
e^{t\cL_g}&=e^{tg^2 Q_0\widetilde \cL_1 Q_0} Q_0+ \ode(e^{-t\Gamma}) + \ode (g)+\ode (g^3 t)\nonumber \\
&=\tau_A\otimes e^{tg^2\Phi_D}\diag_C \tr_{AB} \otimes {\tau_B} + \ode(e^{-t\Gamma}) + \ode (g)+\ode (g^3 t)
\end{align}
where the constants in all $\ode$ are uniform in $t\geq 0$, $g$ small.
\end{cor}
\begin{rem}
0) The identical statements (\ref{id1}) and (\ref{id2}) show that $1/g^2$ is the time scale of the approach to the asymptotic state, as expected.\\
 i)
The full evolution can be approximated by the restriction of $e^{tg^2\tau_A\otimes \Phi_D \otimes \tau_B}$ to $\ran \widetilde Q_0$, 
(provided $\eta$ is larger than the absolute value of the real part of the eigenvalues of $\tau_A\otimes \Phi_D \otimes \tau_B$ in case $H_C=0$). \\
ii) In case $\cL_0=\cD$, we provide in Section \ref{secnodrive} an interpretation of the approximate evolution $e^{tg^2\tau_A\otimes \Phi_D \otimes \tau_B}$ as a classical continuous time Markov process.\\
iii) Set $ F =\max \{|\Re \lambda | \ \lambda \in \sigma(\Phi_D)\}$. When $H_C=0$, the \aj{explicit} term \aj{in (\ref{appdyn})} is \aj{the leading term} if $F<\eta$, and for times which satisfy \aj{$0 \leq  t < \frac{1}{\eps + F}|\ln (g)|/g^2$}, as $g\ra 0$, for any $\eps>0$.
When $H_C\neq 0$, the same is true \aj{for the explicit term in (\ref{appdynhc})}, without constraint on $F$.\\
\aj{iv) This corollary is relevant for an analysis along the lines of \cite{KM}. }
\end{rem}
\begin{proof} Again we prove the statements for $H_C=0$ only, the other case being similar. The first two statements follow from functional calcul, and Proposition \ref{spec2}, taking into account the analyticity of the spectral data involved.
To get the last statement, we observe that since the CPTP map $e^{t\cL_g}$ has a norm which is uniformly bounded in $t\geq 0$ and $g$ small enough, the norm of
\be
 e^{tg^2(\widetilde Q_0\widetilde \cL_1\widetilde Q_0+\ode(g))}\widetilde Q_0(g)=e^{t\cL_g}-\sum_{j\neq k} e^{t\lambda_{jk}(g)}\widetilde Q_{\lambda_{jk}}(g)+\ode(e^{-t\Gamma})
\ee
is bounded above by a constant $C>0$ which uniform in $t\geq 0$ and $g$ small enough. Thus, by Duhamel formula 
\be
e^{\tau (A+B)}=e^{\tau A}+\int_0^\tau e^{\tau'(A+B)}Be^{(\tau-\tau')A}d\tau'
\ee
applied to $A=\widetilde Q_0\widetilde \cL_1\widetilde Q_0$ subject to (\ref{constspec}), $B=\ode(g)$, $\tau = g^2t$ we get
\be
 e^{tg^2(\widetilde Q_0\widetilde \cL_1\widetilde Q_0+\ode(g))}\widetilde Q_0(g)= e^{tg^2\widetilde Q_0\widetilde \cL_1\widetilde Q_0}\widetilde Q_0(g)+\ode (g^3 t).
\ee
Moreover,  $\eta = \min_{j \neq k} \{|\Re \tilde \lambda_{jk}^{(1)}|\}>0$ immediately implies upon expanding $\widetilde Q_0(g)$,
\be
e^{t\cL_g}=e^{tg^2\widetilde Q_0\widetilde \cL_1\widetilde Q_0}\widetilde Q_0+ \ode(e^{-tg^2\eta}) +\ode(g)+\ode (g^3 t), 
\ee
where the constants in all $\ode$ are uniform in $t\geq 0$ and $g$ small.  Finally, \\
$\widetilde Q_0\widetilde \cL_1\widetilde Q_0=\tau_A\otimes \Phi_D\otimes \tau_B$ allows us to express the exponential in terms of that of $\Phi_D$. 
 $\hfill \Box$
\end{proof}

\section{Construction of the asymptotic state}
\label{sec:5}

We now turn to the determination of the state $\rho_0(g)\in \ker \cL_g$ where $\cL_g=\cL_0+g\cL_1\in\cB(\cB(\cH))$ given by a power series in $g$
\be\label{powser}
\rho_0(g)=\rho_0+g\rho_1+g^2\rho_2+\cdots,
\ee
where $\tr (\rho_0)=1$ and $\tr (\rho_j)=0$, $\forall j>0$. 
Expanding $\cL_0(\rho_0(g))+g\cL_1(\rho_0(g))\equiv 0$, and equating like powers of $g$ we get
\begin{align}\label{hierarchy}
\cL_0(\rho_0)&=0 \nonumber \\
\cL_0(\rho_1)&+\cL_1(\rho_0)=0 \nonumber \\
\cL_0(\rho_2)&+\cL_1(\rho_1)=0 \nonumber \\
&\vdots \nonumber \\
\cL_0(\rho_j)&+\cL_1(\rho_{j-1})=0 \ \ \forall \ j\geq 1.
\end{align} 
The way to solve this set of equations, in principle, is as follows. Note that the spectral decomposition of $\cL_0$ yields
\be
\ker  \cL_0=\ran Q_0 \ \ \mbox{and} \ \ \ran \cL_0=\ker Q_0.
\ee
The first equation is solved by picking a trace one element $R_0$ in $\ker \cL_0=Q_0(\cB(\cH))$, described in Proposition \ref{specunc}. The addition of any traceless vector $r_0\in \ker \cL_0$ yields an equally good solution for $\rho_0:=R_0+r_0$ at that order. 
The next equation amounts to solve
$\cL_0(R_1)=-\cL_1(R_0+r_0)$ for $R_1$, a traceless matrix. This requires $\cL_1(R_0+r_0)\in\ran \cL_0$. Since $\ran \cL_0=\ker Q_0$, this is equivalent to 
$Q_0\cL_1Q_0r_0=-Q_0\cL_1R_0$, which determines $r_0=Q_0r_0$ up to the addition of an element of $\ker Q_0\cL_1Q_0$ ($Q_0\cL_1Q_0$ viewed as an operator on $Q_0(\cB(\cH))$). 
Let us assume for the discussion here  that $Q_0\cL_1Q_0\neq 0$, {\it i.e.} $H_C=0$.
This yields $R_1=-\cL_0^{-1}(\cL_1(R_0+r_0))$. Again, the addition of any traceless vector $r_1=Q_0r_1\in \ker \cL_0$ to that $R_1$ yields an equally good solution $\rho_1:=R_1+r_1$ to that equation. 
The next order requires $\cL_1(r_1-\cL_0^{-1}(\cL_1(R_0+r_0)))\in \ran \cL_0$,  which is equivalent to 
$Q_0\cL_1Q_0r_1=Q_0\cL_1\cL_0^{-1}\cL_1(R_0+r_0)$. This equation will then determine $r_0$ completely, under generic hypotheses, as we shall see. Then we proceed by induction.

The case $H_C\neq 0$ is slightly different, see Lemma \ref{redop}, but is approached in the same spirit. 
We start by working out the first few steps and then give the general statements about this construction in Theorem \ref{astat} for $H_C=0$ and Theorem \ref{astat2} for $H_C\neq 0$.
\\

Again, the inverse of $\cL_0$ on its range is the reduced resolvent $S_0=\cL_0^{-1}(\un - Q_0)=\cL_0^{-1}|_{(\un - Q_0)\cB(\cH)}$. To express $S_0$, it is enough to consider the spectral decomposition $\cL_0=\sum_{k>0}\lambda_k Q_k$, where $\lambda_k\neq 0$ and $Q_k$ are the spectral projectors corresponding to Proposition \ref{specunc}, while $\lambda_0=0$ corresponds to the projector $Q_0$.

\subsection{$H_C=0$}

We consider here that $H_C=0$ and work under the spectral assumption {\bf Spec($\overline{H}^{\, \tau}$)}
 on the self-adjoint operator defined by (\ref{hbar}).
We first work out the orders $g^0$ and $g^1$ terms, {\it i.e.} $\rho_0$ and $\rho_1$, and then state an abstract result on the full perturbation series in Theorem \ref{astat}.

\medskip

The first equation yields $R_0=\tau_A\otimes \rho_C \otimes \tau_B$ where $\rho_C$ is a state. We choose $\rho_C=\frac1{n_C}\un_C$, and $r_0=\tau_A\otimes r_C^{(0)}\otimes \tau_B$ with any traceless $r_C^{(0)}$ can be added to that choice so that 
\be\label{leadingHc0}
\rho_0=\tau_A\otimes \rho_C^{(0)}\otimes \tau_B, \ \mbox{with } \ \ \rho_C^{(0)}=\frac1{n_C}\un_C+ r_C^{(0)}.
\ee

Then we compute $Q_0 \cL_1(R_0+r_0)$:
\be
Q_0(-i[H, R_0+r_0 ])=-i\tau_A\otimes [\overline{H}^{\, \tau},  \rho_C^{(0)}]\otimes \tau_B=-i\tau_A\otimes [\overline{H}^{\, \tau}, r_C^{(0)}]\otimes \tau_B.
\ee

The condition to solve the equation for $R_1$ requires $r_C^{(0)}=\diag_\tau(r_C^{(0)})$, where $\diag_\tau(\cdot)$ extracts the diagonal part of $r_C^{(0)}$ in the normalised eigenbasis of $\overline{H}^{\, \tau}$. Thanks to our assumption, we set
\be\label{defrho1}
R_1:=i\cL_0^{-1}([H, R_0+r_0])=i\sum_{k>0}\lambda_k^{-1} Q_k([H, R_0+r_0]),
\ee
which is traceless, since $R_1=(\un-Q_0)R_1$, and self-adjoint if $r_C^{(0)}$ is. Next we look for $R_2$, which requires $Q_0(\cL_1(R_1+r_1))=0$, where $r_1=Q_0(r_1)=\tau_A\otimes r_C^{(1)}\otimes \tau_B$:
\be
Q_0([H,\{\cL_0^{-1}(i[H, \tau_A\otimes \diag_\tau(\rho_C^{(0)}) \otimes \tau_B])+\tau_A\otimes r_C^{(1)}\otimes \tau_B\}])=0.
\ee
This is equivalent to the equation on $\cB(\cH_C)$
\be\label{diagr}
i\tr_{AB}\big(\big[H, \cL_0^{-1}([H, \tau_A\otimes \diag_\tau (\rho_C^{(0)}) \otimes \tau_B])\big]\big)+ [\overline{H}^{\, \tau}, r_C^{(1)}]=0,
\ee
where we note that $\diag_\tau (r_C^{(1)})$ is arbitrary.
Our hypotheses on $\overline{H}^{\, \tau}$ imply that
\begin{align}
\ker [\overline{H}^{\, \tau},  \cdot  ]&=\{\rho_C\, | \, \rho_C=\diag_\tau \rho_C \}, \\
\ran  [\overline{H}^{\, \tau},  \cdot  ]&=\{\rho_C\, | \, \rho_C=\offdiag_\tau \rho_C\}.
\end{align}
Now, assumption {\bf Coup } on  $H$ ensures (\ref{diagr}) determines $\diag_\tau r_C^{(0)}$ and $\offdiag_\tau r_C^{(1)}$:
Separating the diagonal from the offdiagonal parts, we have for the former
\begin{align}
\Phi_D( \rho_C^{(0)})=0,
\end{align}
which determines $\rho_C^{(0)}=\un_C/n_C+\diag_\tau r_C^{(0)}=\diag_\tau (\rho_C^{(0)})$ fully since $\dim \ker \Phi_D=1$, and thus  $R_1$ as well. The offdiagonal part yields
\begin{align}\label{offdiagrc1}
\offdiag_\tau r_C^{(1)}&= -i [\overline{H}^{\, \tau},  \cdot  ]^{-1}\Big( \offdiag_\tau \tr_{AB}\big(\big[H, \cL_0^{-1}([H, \tau_A\otimes \rho_C^{(0)} \otimes \tau_B])\big]\big)\Big)\nonumber \\
&=-i[\overline{H}^{\, \tau},  \cdot  ]^{-1}\Big( \offdiag_\tau \Phi (\rho_C^{(0)})) \Big)
\end{align}
which fixes $\offdiag_\tau r_C^{(1)}$ and leaves $\diag_\tau r_C^{(1)}$ open for now.

At this point, the formula which defines $R_2$  makes sense,
\be\label{defrho2}
R_2=i\cL_0^{-1}([H, R_1+r_1])=i\sum_{k>0}\lambda_k^{-1} Q_k([H, R_1+r_1]),
\ee
where $R_2$ depends parametrically on $\diag_\tau r_C^{(1)}$. At order two, the contribution is $R_2+r_2$, where $r_2=Q_0(r_2)=\tau_A\otimes r_C^{(2)}\otimes \tau_B$ is arbitrary.
The term $\diag_\tau r_C^{(1)}$ is determined by the requirement that $Q_0(\cL_1(R_2+r_2))=0$ necessary to solve for $R_3$, {\it i.e.}
\begin{align}
\tr_{AB}([H,&\{\cL_0^{-1}(i[H, R_1+\tau_A\otimes r_C^{(1)}\otimes \tau_B])+\tau_A\otimes r_C^{(2)}\otimes \tau_B\}])\\ \nonumber
&=\tr_{AB}([H,\cL_0^{-1}(i[H, R_1+\tau_A\otimes r_C^{(1)}\otimes \tau_B])+[\overline{H}^{\, \tau}, r_C^{(2)}]=0.
\end{align}
Splitting this equation into its diagonal and offdiagonal parts, we get, making use of (\ref{defphi}),
\begin{align}
&\diag_\tau \tr_{AB}([H,\cL_0^{-1}(i[H, R_1+\tau_A\otimes \offdiag_\tau r_C^{(1)}\otimes \tau_B])]+ \Phi_D (\diag_\tau r_C^{(1)})=0, \\
&\offdiag_\tau \tr_{AB}([H,\cL_0^{-1}(i[H, R_1+ \tau_A\otimes r_C^{(1)}\otimes \tau_B])]+[\overline{H}^{\, \tau}, r_C^{(2)}]=0.
\end{align}
Using assumption {\bf Coup} under the form:
$\Phi_D$
 is invertible on the subspace 
$ \ran \Phi_D=\diag_\tau \cB(\cH_C)\cap \{\rho_C\, | \, \tr \rho_C=~0\}$, the first equation determines 
\be\label{diagrc1}
\diag_\tau r_C^{(1)}=-\Phi_D^{-1}(\diag_\tau \tr_{AB}([H,\cL_0^{-1}(i[H, R_1+\tau_A\otimes\offdiag_\tau r_C^{(1)}\otimes\tau_B])]),
\ee 
so that $r_C^{(1)}$ is determined and therefore the second equation yields
\be
\offdiag_\tau r_C^{(2)}=-i[\overline{H}^{\, \tau},  \cdot  ]^{-1}\Big( \offdiag_\tau \tr_{AB}\big(\big[H, \cL_0^{-1}([H, R_1+\tau_A\otimes r_C^{(1)} \otimes \tau_B])\big]\big)\Big).
\ee
Consequently, we can set
\be
R_3=i\cL_0^{-1}([H, R_2+r_2])=(\un-Q_0)R_3.
\ee
At this point, $\rho_0=R_0+r_0$, $\rho_1=R_1+r_1$ are known, as well as $R_2$, $\offdiag_\tau r_C^{(2)}$ and $R_3$.

\begin{rem} \label{zerotrace} 
The fact that $\rho_C^{(0)}\in \ker \Phi_D$ implies $\tr \rho_C^{(0)}\neq 0$, so the assumption that  $\rho_C$ is a state in the initial step amounts to set a normalisation.
\end{rem}

Let us formulate a general result that summarises the foregoing and guarantees the process can be pursued:
\begin{thm}\label{astat}
Consider the \qrm Lindbladian $\cL_g$ (\ref{lgthm})  with $H_C=0$ under the assumptions  {\bf Spec($\overline{H}^{\, \tau}$)}
and {\bf Coup}. 
Then there exists $g_0>0$ such that  $\rho_0(g)$, the unique invariant state of $\cL_g$, admits a convergent  expansion
\be
\rho_0(g)=\rho_0+g\rho_1+g^2\rho_2+\cdots,
\ee
for all $g\in \C$ with $|g|<g_0$. We have, 
\begin{align}
\rho_0&=\tau_A\otimes \rho_C^{(0)} \otimes \tau_B, \ \mbox{where} \ \rho_C^{(0)} \in \ker \Phi_D
\end{align}
see (\ref{defphi}) and (\ref{leadingHc0}), and 
\be
\rho_j=R_j+\tau_A\otimes r_C^{(j)}\otimes \tau_B
\ee
 for all $j\geq 1$.  
Moreover, there exists a linear map  $\cR: \cB(\cH)\ra  \cB(\cH)\cap \{\rho_C\, | \, \tr \rho_C=~0\}$ such that
$
\rho_j=\cR(\rho_{j-1}),
$
where
\begin{align}
&R_j=i\cL_0^{-1}([H, \rho_{j-1}]),
\\
&\offdiag_\tau r_C^{(j)}=-i[\overline{H}^{\, \tau},  \cdot  ]^{-1}\Big( \offdiag_\tau \tr_{AB}\big(\big[H, \cL_0^{-1}([H, \rho_{j-1}])\big]\big)\Big),
\\
&\diag_\tau r_C^{(j)}=-\Phi_D^{-1}\big(\diag_\tau \tr_{AB}([H,\cL_0^{-1}(i[H, R_{j}+\tau_A\otimes\offdiag_\tau r_C^{(j)}\otimes\tau_B])])\big).\label{offdiagrj}
\end{align}
{  Consequently, for $|g|<g_0$,
\be\label{rgresolv}
\rho_0(g)=(\un-g\cR)^{-1}(\rho_0).
\ee
}
\end{thm}
\begin{rem}\label{strucR}
0) Replacing $R_j$ and $\offdiag_\tau r_C^{(j)}$ by their expression into (\ref{offdiagrj}) shows $\diag_\tau r_C^{(j)}$ is linear in  $\rho_{j-1}$ as well an yields the map $\cR$.\\
i) Eq. (\ref{rgresolv}) is equivalent to 
\be\label{resolvR}
\rho_0(g)=\sum_{k=1}^{N} \left(\frac{M_k}{1-g\mu_k}+\sum_{l=1}^{m_k-1}\frac{g^l N_k^l}{(1-g\mu_k)^{l+1}}\right)(\rho_0),
\ee
where $\mu_k, M_k, N_k$ and $m_k$ are the eigenvalues, eigenprojectors, eigennilpotents and algebraic multiplicities appearing in the spectral decomposition of $\cR=\sum_{k=1}^N\mu_k M_k+N_k$. Hence the radius of convergence is $g_0=1/\max_{1\leq k\leq N}(|\mu_k|)$.  \\
ii) In case $\sigma(\cR)\cap \R_\pm^*=\emptyset$, the steady state $\rho_0(g)$ is well defined for all $g\in \R_\pm^*$.\\
iii) The iteration terminates if and only if $\cR$ has a zero eigenvalue and $\rho_0$ belongs to the corresponding eigenspace; see Section \ref{sec:example2} for examples. \\
iv) The restriction of the invariant state to $\cH_C$ is given by $\tr_{AB}(\rho_0(g))=\rho_C^{(0)}+\sum_{j\geq 1}g^jr_C^{(j)}$.\\
v)  We provide necessary and sufficient conditions in Proposition \ref{couphold} for {\bf Coup} to be satisfied in case $\cL_0=\cD$ and $H_C=0$.

\end{rem}
\begin{proof}
Recall that $\dim \ker \cL_g=1$ is proven in Theorem \ref{goldenrule}.\\
We  solve the higher orders equations for $\rho_j=R_j+r_j$ of (\ref{hierarchy}) with  
\be \label{rq}
R_j=(\un-Q_0)R_j, \  r_{j}=Q_0 r_{j}=\tau_A\otimes r_C^{(j)}\otimes \tau_B,
\ee
for all $j$ by induction. 
Let $j\geq 2$ and assume $R_{k}$, $r_{k}$ are given traceless matrices satisfying (\ref{rq}) for $1\leq k\leq j-1$ as well as
\begin{align}
&R_{j}=i\cL_0^{-1}([H,R_{j-1}+r_{j-1}]), \  \tau_A\otimes\offdiag_\tau r_C^{(j)}\otimes \tau_B\ \mbox{and} \
R_{j+1}=i\cL_0^{-1}([H,R_{j}+r_{j}]).
\end{align} 
This is the situation we arrived at for $j=2$.
Consider $Q_0(\cL_1(R_{j+1}+r_{j+1}))=0$, a necessary condition to compute
$R_{j+2}$, which yields
\begin{align}
\tr_{AB}\big([H,&\cL_0^{-1}(i[H, R_{j}+\tau_A\otimes r_C^{(j)}\otimes \tau_B])\big)+[\overline{H}^{\, \tau}, r_C^{(j)}]=0.
\end{align}
Splitting the equation into its diagonal and offdiagonal parts gives
\begin{align}
&\diag_\tau \tr_{AB}\big([H,\cL_0^{-1}(i[H, R_{j}+\tau_A\otimes \offdiag_\tau r_C^{(j)}\otimes \tau_B]\big)+\Phi_D(\diag_\tau r_C^{(j)})=0, \\
&\offdiag_\tau \tr_{AB}\big([H,\cL_0^{-1}(i[H, R_{j}+\tau_A\otimes r_C^{(j)}\otimes \tau_B]\big)+[\overline{H}^{\, \tau}, r_C^{(j+1)}]=0.
\end{align}
The first equation determines 
\be
\diag_\tau r_C^{(j)}=-\Phi_D^{-1}\big(\diag_\tau \tr_{AB}([H,\cL_0^{-1}(i[H, R_{j}+\tau_A\otimes\offdiag_\tau r_C^{(j)}\otimes\tau_B])])\big),
\ee 
so that $r_C^{(j)}$ is fully determined and therefore the second equation yields
\be
\offdiag_\tau r_C^{(j+1)}=-i[\overline{H}^{\, \tau},  \cdot  ]^{-1}\Big( \offdiag_\tau \tr_{AB}\big(\big[H, \cL_0^{-1}([H, R_{j}+\tau_A\otimes r_C^{(j)} \otimes \tau_B])\big]\big)\Big).
\ee
Consequently we can define
\be
R_{j+2}=i\cL_0^{-1}([H,R_{j+1}+r_{j+1}]),
\ee
where $\diag_\tau r_C^{(j+1)}$ remains free, while $r_j$ is determined. This finishes the proof of the induction.
\end{proof} $\hfill \Box$

\subsection{$H_C\neq$ 0}

We consider here $H_C\neq 0$ and the necessary modifications to compute the series (\ref{powser}) due to the identities
\be\label{idzero}
Q_0 (\cdot)=\tau_A\otimes \diag_C(\tr_{AB}(\cdot ))\otimes \tau_B \ \, \mbox{and} \ \, Q_0\cL_1Q_0\equiv 0.
\ee 
The first equation in (\ref{hierarchy}) yields $\rho_0=Q_0 \rho_0=\tau_A\otimes \rho_C^{(0)} \otimes \tau_B$, where $\rho_C^{(0)}~\in~\diag_C\cB(\cH_C)$ is free.
The condition to solve the second equation is 
$Q_0\cL_1(\rho_0)=Q_0\cL_1Q_0(\rho_0)=0$ which is trivially satisfied. Thus, writing $\rho_1=R_1+r_1$ with $R_1=(\un-Q_0)\rho_1$ and $r_1=Q_0 \rho_1$, we can solve partially the equation setting 
\be
R_1=-\cL_0^{-1}\cL_1(\rho_0).
\ee 
The next equation $\cL_0(\rho_2)=-\cL_1(\rho_1)$ requires $Q_0\cL_1(R_1)+Q_0\cL_1 (r_1)=0$. Thanks to $r_1=Q_0 r_1$ and the identity (\ref{idzero}), this equation reduces to 
\be \label{eqrho0}
Q_0\cL_1\cL_0^{-1}\cL_1 Q_0(\rho_0)=0,
\ee
 where we used the expression for $R_1$ and $\rho_0=Q_0 \rho_0$. Thanks to assumption {\bf Coup} for $H_C\neq 0$, this determines $\rho_0=\tau_A\otimes \diag_C\rho_C^{(0)} \otimes \tau_B $ since (\ref{eqrho0}) is equivalent to
\be\label{rho0ker}
 \rho_C^{(0)} \in \ker \Phi_D, \ \mbox{where} \ \dim \ker \Phi_D=1.
\ee
Thus $R_1$ is now determined, while the traceless part $r_1=\tau_A\otimes \diag_C r_C^{(1)}\otimes \tau_B$ is not. With the familiar decomposition $\rho_2=R_2+r_2$ with respect to the projector $Q_0$, we set
\be\label{eqR2}
R_2=-\cL_0^{-1}\cL_1(R_1+r_1)
\ee
and turn to the equation for $\rho_3=R_3+r_3$: $\cL_0(\rho_3)=\cL_0(R_3)=-\cL_1(\rho_2)$. It requires $Q_0\cL_1(R_2+r_2)=Q_0\cL_1(R_2)=0$, where we used (\ref{idzero}) and $r_2=Q_0 r_2$. With (\ref{eqR2}), this is equivalent to
\be
Q_0\cL_1\cL_0^{-1}\cL_1 Q_0(r_1)=-Q_0\cL_1\cL_0^{-1}\cL_1(R_1)=-\tau_A\otimes \diag_C \tr_{AB}(\cL_1\cL_0^{-1}\cL_1(R_1))\otimes \tau_B,
\ee
where $\tr \cL_1\cL_0^{-1}\cL_1(R_1)=0$, since $\cL_1(\cdot)=-i[H, \cdot ]$. Thanks to {\bf Coup}, we can thus determine $r_1=\tau_A\otimes \diag_C r_C^{(1)}\otimes \tau_B$ uniquely in terms of $\Phi_D$
\be
r_C^{(1)}=\Phi_D^{-1}\Big(\diag_C \tr_{AB}\big\{\big[H,\cL_0^{-1}([H, R_1]\big]\big\} \Big).
\ee
In turn $R_2$ is fully determined while $r_2=\tau_A\otimes \diag_C r_C^{(2)}\otimes \tau_B$ remains to be computed, and 
\be
R_3=-\cL_0^{-1}\cL_1(R_2+r_2).
\ee
From there on we can iterate the process to get the equivalent of Theorem \ref{astat} in the case  $H_C\neq 0$. The proof being similar and simpler, we omit it.
\begin{thm}\label{astat2}
Consider the \qrm Lindbladian $\cL_g$ (\ref{lgthm})  with $H_C\neq 0$ under the assumptions  {\bf Spec($\overline{H}^{\, \tau}$)}
and {\bf Coup}. 
Then there exists $g_0>0$ such that  $\rho_0(g)$, the unique invariant state of $\cL_g$, admits a convergent  expansion
\be
\rho_0(g)=\rho_0+g\rho_1+g^2\rho_2+\cdots,
\ee
for all $g\in \C$ with $|g|<g_0$. We have, 
\begin{align}
\rho_0&=\tau_A\otimes \rho_C^{(0)} \otimes \tau_B, \ \mbox{where} \ \rho_C^{(0)} \in \ker \Phi_D
\end{align}
see (\ref{defphi}) and (\ref{rho0ker}), and $\rho_j=R_j+\tau_A\otimes r_C^{(j)}\otimes \tau_B$ for all $j\geq 1$, with $ r_C^{(j)}=\diag_C ( r_C^{(j)})$.  
Moreover, there exists a linear map  $\cR: \cB(\cH)\ra  \cB(\cH)\cap \{\rho_C\, | \, \tr \rho_C=0\}$ such that
$
\rho_j=\cR(\rho_{j-1}),
$
where
\begin{align}\label{iteratehc}
R_j&=i\cL_0^{-1}([H, \rho_{j-1}]),\\
 r_C^{(j)}&=\Phi_D^{-1}\big(\diag_C \tr_{AB}([H,\cL_0^{-1}([H, R_{j}])])\big)\nonumber \\
&=i\Phi_D^{-1}\big(\diag_C \tr_{AB}([H,\cL_0^{-1}([H, \cL_0^{-1}([H, \rho_{j-1}])])])\big).
\end{align}
{  Consequently, for $|g|<g_0$,
\be
\rho_0(g)=(\un-g\cR)^{-1}(\rho_0).
\ee
}
\end{thm}
\begin{rem}
0) Remarks i), ii, iii) below Theorem \ref{astat} remain in force here.\\
i) The map $\cR$ can be expressed as 
\begin{align}
R_{j}&=-\cL_0^{-1}\cL_1(\rho_{j-1}), \\ \nonumber
 r_C^{(j)}&=-\Phi_D^{-1}\big(\tr_{AB}\big\{ Q_0\cL_1\cL_0^{-1}\cL_1 (R_{j})\big\}\big)=\Phi_D^{-1}\big(\tr_{AB}\big\{ Q_0\cL_1\cL_0^{-1}\cL_1 \cL_0^{-1}\cL_1(\rho_{j-1})\big\}\big)
\end{align}
so that
\be
\rho_j=\Big(-\cL_0^{-1}\cL_1(\cdot)+\tau_A\otimes \Phi_D^{-1}\big(\tr_{AB}\big\{ Q_0\cL_1\cL_0^{-1}\cL_1 \cL_0^{-1}\cL_1(\cdot)\big\}\big) \otimes \tau_B\Big)(\rho_{j-1}).
\ee
\end{rem}

\section{
No leading order Hamiltonian drive
}\label{secnodrive}

We consider here the case where $H_A=H_B=H_C=0$ on their respective spaces, so that $\cL_0=\cD$ with $\tau_A$ and $\tau_B$ arbitrary, while $\cL_1=-i[H,\, \cdot \, ]$ with $H$ arbitrary as well. This allows us to keep things relatively simple, while retaining a certain level of generality, since the dimensions of the different Hilbert spaces are arbitrary as well.

\medskip

Let us consider the hypothesis {\bf Coup} in this simplified setup, assuming {\bf Spec($\overline{H}^{\, \tau}$)} holds. Recall that $\{\ffi_j^\tau\}_{1\leq j\leq n_C}$ denotes the normalized eigenbasis of $\overline{H}^{\, \tau}$ with respect to which the projectors $\diag_\tau$ and $\offdiag_\tau$ are defined, and set $P_j^\tau=|\ffi_j^\tau\ket\bra \ffi_j^\tau|$. Given the definition (\ref{defphi}) of $\Phi_D$, we need to compute for all $j,k\in\{1,\dots, n_C\}$
\be
(\Phi_D)_{jk}:=\tr \big\{(\un_A\otimes P_j^\tau\otimes \un_B)\big([H,\cL_0^{-1}( [H,\tau_A\otimes P_k^\tau\otimes \tau_B]\big)\big\}.
\ee
Thanks to Proposition \ref{specdiss}, we can express $\cL_0^{-1}=\cD^{-1}$ in a compact way. Let $\tilde \rho_0\in \cB(\cH)$ such that $\tr_{AB}(\tilde \rho_0)=0$, so that $Q_0(\tilde \rho_0)=0$. Thus
\begin{align}
\cD^{-1}(\tilde \rho_0)=\frac{-1}{\gamma_A+\gamma_B}\left\{  \tilde \rho_0+\frac{\gamma_A}{\gamma_B} \tau_A \otimes \tr_A(\tilde \rho_0)+\frac{\gamma_B}{\gamma_A}\tr_B(\tilde \rho_0)\otimes \tau_B\right\}
\end{align}
Therefore, introducing
\begin{align}
\overline{H}^{\, \tau_A}&=\tr_{A}(H(\tau_A\otimes \un_C \otimes \un_B))=\tr_{A}((\tau_A\otimes \un_C \otimes \un_B)H)\in \cB(\cH_C\otimes \cH_B),\\
\overline{H}^{\, \tau_B}&=\tr_{B}(H(\un_A\otimes \un_C \otimes \tau_B))=\tr_{B}((\un_A\otimes \un_C \otimes \tau_B)H)\in \cB(\cH_A\otimes \cH_C)
\end{align}
and making use of $\tr_{AB} [H,\tau_A\otimes P_k^\tau\otimes \tau_B]=0$, a straightforward computation yields

\begin{align}\label{hlm1}
[H,\cL_0^{-1}( [H,&\tau_A\otimes P_k^\tau\otimes \tau_B])]=-\frac{1}{\gamma_A+\gamma_B} [H,[H,\tau_A\otimes P_k^\tau\otimes \tau_B]]
\\ \nonumber
&-\frac{\gamma_A/\gamma_B}{\gamma_A+\gamma_B}[H,\tau_A\otimes [\overline{H}^{\, \tau_A}, P_k^\tau\otimes\tau_B]]-\frac{\gamma_B/\gamma_A}{\gamma_A+\gamma_B}[H,[\overline{H}^{\, \tau_B},\tau_A\otimes P_k^\tau]\otimes \tau_B].
\end{align}

Then we note using the cyclicity of the trace that
\begin{align}
\tr \big\{(\un_A\otimes &P_j^\tau\otimes \un_B)\big([H,[H,\tau_A\otimes P_k^\tau\otimes \tau_B]])\big\} \\ \nonumber
&=2\big(\delta_{jk}\tr (H(\tau_A\otimes P_k^\tau\otimes \tau_B)H)
-\tr ((\un_A\otimes P_j^\tau\otimes \un_B)H(\tau_A\otimes P_k^\tau\otimes \tau_B)H)\big)
\end{align}
where the operator in the first trace reads 

\be
\big((\tau_A^{1/2}\otimes P_k^\tau\otimes \tau_B^{1/2})H\big)^*(\tau_A^{1/2}\otimes P_k^\tau\otimes \tau_B^{1/2})H\geq 0,
\ee
while the second trace yields the $jj$ element of its partial $\tr_{AB}$. Hence,
\begin{align}
\tr \big\{(\un_A\otimes P_j^\tau\otimes \un_B)\big([H,&[H,\tau_A\otimes P_k^\tau\otimes \tau_B]])\big\}\\ \nonumber 
&=2
\left\{ \begin{matrix} 
 - \tr_{AB} (H(\tau_A\otimes P_k^\tau\otimes \tau_B)H)_{jj}\leq 0& \mbox{if} \ j\neq k \cr 
    \sum_{l\neq k}\tr_{AB} (H(\tau_A\otimes P_k^\tau\otimes \tau_B)H)_{ll}\geq 0& \mbox{if} \ j = k
\end{matrix}\right. .
\end{align}
Similar considerations can be made for the traces of the other two operators in (\ref{hlm1}):
\begin{align}
\tr \big\{(\un_A\otimes P_j^\tau &\otimes \un_B)\big([H,[\overline{H}^{\, \tau_B},\tau_A\otimes P_k^\tau]\otimes \tau_B]\big)\big\} =
\tr \big\{(\un_A\otimes P_j^\tau)\big([\overline{H}^{\, \tau_B},[\overline{H}^{\, \tau_B},\tau_A\otimes P_k^\tau]])\big\} \nonumber \\ \nonumber
&=2\big(\delta_{jk}\tr (\overline{H}^{\, \tau_B}(\tau_A\otimes P_k^\tau)\overline{H}^{\, \tau_B})
-\tr ((\un_A\otimes P_j^\tau)\overline{H}^{\, \tau_B}(\tau_A\otimes P_k^\tau)\overline{H}^{\, \tau_B})\big)\\ 
&=2
\left\{ \begin{matrix} 
 - \tr_{A} (\overline{H}^{\, \tau_B}(\tau_A\otimes P_k^\tau)\overline{H}^{\, \tau_B})_{jj}\leq 0& \mbox{if} \ j\neq k \cr 
    \sum_{l\neq k}\tr_{A} (\overline{H}^{\, \tau_B}(\tau_A\otimes P_k^\tau)\overline{H}^{\, \tau_B})_{ll}\geq 0& \mbox{if} \ j = k
\end{matrix}\right. ,
\end{align}
and
\begin{align}
\tr \big\{(\un_A\otimes P_j^\tau &\otimes \un_B)\big([H,\tau_A\otimes [\overline{H}^{\, \tau_A},P_k^\tau \otimes \tau_B]]\big)\big\} =
\tr \big\{(P_j^\tau \otimes \un_B)\big([\overline{H}^{\, \tau_A},[\overline{H}^{\, \tau_A},P_k^\tau \otimes \tau_B]])\big\} \nonumber \\ \nonumber
&=2\big(\delta_{jk}\tr (\overline{H}^{\, \tau_A}(P_k^\tau \otimes \tau_B)\overline{H}^{\, \tau_A})
-\tr ((P_j^\tau \otimes \un_A)\overline{H}^{\, \tau_A}(P_k^\tau \otimes \tau_B)\overline{H}^{\, \tau_A})\big)\\ 
&=2
\left\{ \begin{matrix} 
 - \tr_{B} (\overline{H}^{\, \tau_A}(P_k^\tau \otimes \tau_B)\overline{H}^{\, \tau_A})_{jj}\leq 0& \mbox{if} \ j\neq k \cr 
    \sum_{l\neq k}\tr_{B} (\overline{H}^{\, \tau_B}(P_k^\tau \otimes \tau_B)\overline{H}^{\, \tau_B})_{ll}\geq 0& \mbox{if} \ j = k
\end{matrix}\right. .
\end{align}

Defining for $1\leq k \leq n_C$ the non negative operator $h(k)\in\cB(\cH_C)$ by 
\begin{align}\label{posh}
h(k)=&\frac{2}{\gamma_A+\gamma_B}\tr_{AB} (H(\tau_A\otimes P_k^\tau\otimes \tau_B)H)\\ \nonumber
&+\frac{2\gamma_A/\gamma_B}{\gamma_A+\gamma_B}\tr_{B} (\overline{H}^{\, \tau_A}(P_k^\tau \otimes \tau_B)\overline{H}^{\, \tau_A}) +\frac{2\gamma_B/\gamma_A}{\gamma_A+\gamma_B}\tr_{A} (\overline{H}^{\, \tau_B}(\tau_A\otimes P_k^\tau)\overline{H}^{\, \tau_B}),
\end{align}
we eventually obtain
\be\label{transmat}
(\Phi_D)_{jk}=
\left\{ \begin{matrix} 
  - h(k)_{jj}&\geq 0& \mbox{if} \ j\neq k \cr 
    +\sum_{l\neq k} h(k)_{ll}&\neq 0& \mbox{if} \ j = k
\end{matrix}\right. ,
\ee
where $\Phi_D$ is viewed as a matrix on $\C^{n_C}$, and any diagonal matrix $r=\sum_{k=1}^{n_C}{r_k}P_k^\tau\in \diag_\tau \cB(\cH_C)$ is viewed as a vector $\begin{pmatrix} r_1 &r_2 &\cdots &r_{n_C}\end{pmatrix}^t$ of $\C^{n_C}$.
\medskip

We provide a necessary and sufficient condition on the coupling Hamiltonian $H$ in terms of the diagonal matrix elements of $h(k)$, $1\leq k\leq n_C$ for  assumption {\bf Coup} to hold, {\it i.e.} that $\Phi_D$ restricted to diagonal traceless matrices is invertible.

\begin{prop}\label{couphold}
Assume $\cL_0=\cD$, $\cL_1=-i[H,\cdot \, ]$ and consider the non negative operators $\{h(k)\}_{1\leq k\leq n_C}$ defined by (\ref{posh}). Assumption  {\bf Coup} holds if and only if 
there exists $j\in \{1, \dots, n_C\}$ such that 
$h_{jj}(k)>0$ for all $1\leq k\neq j \leq n_C$.
\end{prop}
\begin{rem}
i) Since $h(k)$ is a sum of non negative operators, it is sufficient to check the condition on any of its constituants. \\
ii) Explicit computations show that for $\dim \cH_C=2$, assumption {\bf Coup} holds as soon as $\Phi_D\neq 0$, while for $\dim \cH_C=3$ it is true if
$h_r(j)h_s(k)>0$ for some $1\leq j\neq k\leq 3$, $r\neq j$, $s\neq k$ and $(r,s)\neq (k,j)$.
\end{rem}

\begin{proof}
Within the framework introduced above we identify  $\Phi_D$ with its matrix $(\Phi_D)_{jk}$. We need to show it admits zero as a simple eigenvalue, which amounts to showing that $\rank  \Phi_D=n_C-1$.

We use the short hand notations $h_j(k)=h(k)_{jj}\geq 0$ for $j\neq k$ and $h_j(j)=\sum_{k\neq j}h_k(j)\geq 0$ to express the matrix elements of $-\Phi_D$. The proof  follows once we establish the following Lemma
\begin{lem}\label{math} Consider ${\mathfrak h}\in M_n(\R)$ given by
\begin{align}\label{matphid}
{\mathfrak h} = \begin{pmatrix}
h_1(1) & -h_1(2) & \cdots & -h_1(n) \cr
-h_2(1) & h_2(2) &  & -h_2(n) \cr
\vdots & &\ddots & \vdots \cr
-h_n(1) & -h_n(2) & \cdots & h_n(n)
\end{pmatrix}, \ \mbox{where }\ \left\{\begin{matrix}h_j(k)\geq 0 \ \ \ \ \  \, \mbox{for } \  j\neq k \cr h_j(j)=\sum_{k\neq j}h_k(j) \geq 0 \end{matrix}\right. .
\end{align}
Then, $\rank \, {\mathfrak h}=n-1$ if and only if $\ \exists \ 1\leq j\leq n$ such that $h_j(k)>0$,  $\forall\  1\leq k \neq j \leq n$.
\end{lem}
\begin{rem}
It is possible that $\rank \, {\mathfrak h}=n-1$ and one diagonal element $h_j(j)=0$, in which case ${\mathfrak h}e_j=0$, where $e_j$ is the $j^{\mbox{\scriptsize th}}$ canonical basis vector of $\C^n$.\\
We can associate to ${\mathfrak h}$ a stochastic matrix ${\mathfrak p}$ the elements of which are
\be
{\mathfrak p}_{jk}=\left\{\begin{matrix} \frac{h_k(j)}{2h_j(j)} & \mbox{\rm if } \ h_j(j)>0 \cr \delta_{jk} & \mbox{\rm if } \ h_j(j)=0\end{matrix} \right. ,
\ee
such that  $x\in\C^n$ satisfies ${\mathfrak h}x=0$ iff ${\mathfrak p}^ty=y$, where $y=\diag({\mathfrak h})x\in \C^n$, if $h_k(k)>0$ for all $k$. Hence, if $\rank \, {\mathfrak h}=n-1$, the components of $x$ can all be chosen to be non negative, by Perron Frobenius theorem. \\
However $\mathfrak p$ is not necessarily irreducible as one sees from the example 
${\mathfrak h} = \begin{pmatrix}
1 & 0 & 0 \cr
-1 & 1 &  -1 \cr
0& -1 & 1
\end{pmatrix}$ with $\sigma({\mathfrak h} )=\{0,1,2\}$ that admits the non strictly positive eigenvector $\begin{pmatrix} 0 & 1 & 1 \end{pmatrix}^T$ in its kernel

\end{rem}

\begin{proof}
We know $0\in\sigma({\mathfrak h})$ and by Jacobi's formula, 
\be
\frac{d}{dz}\det ({\mathfrak h}-z)|_{z=0}=\tr \ \mbox{com}^t({\mathfrak h})=\sum_{j=1}^n\det \hat {\mathfrak h}_{jj},
\ee
 where $ \mbox{com}(A)$ is the comatrix of $A$ and $\hat A_{jk}$ is obtained by deleting the $j^{\mbox \scriptsize \rm th}$ row and $k^{\mbox \scriptsize \rm th}$ column
of  $A$. In our case
\begin{align}\label{minor}
\hat {\mathfrak h}_{jj}=\begin{pmatrix}
h_1(1) &\cdots & -h_1(j-1) & -h_1(j+1) & \cdots & -h_1(n) \cr
\vdots & \ddots& &\vdots & & \vdots \cr
-h_{j-1}(1) & \cdots & h_{j-1}(j-1) & -h_{j-1}(j+1) & &-h_{j-1}(n) \cr
-h_{j+1}(1)  & & -h_{j+1}(j-1)& h_{j+1}(j+1)& & -h_{j+1}(n) \cr
\vdots & & \vdots & &\ddots & \vdots \cr
-h_n(1) & \cdots &-h_n(j-1) & -h_n(j+1) &\cdots & h_n(n)
\end{pmatrix}
\end{align}
is real valued so that $\sigma (\hat {\mathfrak h}_{jj})=\overline{\sigma (\hat {\mathfrak h}_{jj})}$. Moreover, by definition, for all $k\neq j$
\be
h_k(k)=\sum_{l\neq k}h_l(k)\geq \sum_{l\neq k \atop l\neq j}h_l(k),
\ee
so that by Gershgorin Theorem
\be\label{gersh}
\sigma (\hat {\mathfrak h}_{jj})\subset \bigcup_{k\neq j}\Big\{z\in \C \, | \, |z-h_k(k)|\leq  \sum_{l\neq k\atop l\neq j} h_l(k)\Big\}\equiv  \bigcup_{k\neq j} G_k
\ee
where the circle $G_k$ centered at $h_k(k)$ of radius $\sum_{l\neq k\atop l\neq j} h_l(k)$  intersects the imaginary axis if and only if $h_j(k)=0$, in which case the intersection reduces to the origin. Since the determinant of $ \hat {\mathfrak h}_{jj}$ is the product of its complex conjugate eigenvalues, (\ref{gersh}) yields
\be
\det \hat {\mathfrak h}_{jj}\geq 0, \ \mbox{with equality iff} \ \exists \  k\neq j \ \mbox{s.t.} \ h_j(k)=0.
\ee
Therefore
\be
\sum_{j=1}^n \det \hat {\mathfrak h}_{jj}\geq 0,  \ \mbox{with equality iff} \ \forall \ 1\leq j\leq n, \ \exists \  k\neq j \ \mbox{s.t.} \ h_j(k)=0.
\ee
\end{proof}$\hfill \Box$\\
This ends the proof of the Proposition.
\end{proof}$\hfill \Box$

\subsection{Emergence of a classical Markov process}

Coming back to Corollary \ref{dynimp}, we know that for times s.t. \aj{$0\leq t\leq \frac{1}{F+\eps} |\ln(g)|/g^2$}, the evolution semigroup $e^{t(\cD(\cdot)-ig[H,\cdot])}$ can be approximated by 
\be
e^{tg^2\Phi_D}: \diag_\tau \cB(\cH_C)\ra \diag_\tau \cB(\cH_C).
\ee
In the case at hand, $\Phi_D$ is expressed  in the orthonormal basis $\{ |\ffi_j^\tau\ket\bra \ffi_j^\tau |\}_{1\leq j\leq n_C}$ as the matrix (\ref{transmat}) denoted by $\mathfrak h$ in Lemma \ref{math}. The negative of the transpose ${\mathfrak h}^T$ of $\mathfrak h$ is thus a {\it transition rate matrix} or {\it Q-matrix}, associated to a classical continuous time Markov chain with finitely many states, see \cite{N}. 
Therefore we can associate to our quantum problem $\dot \rho = \cD(\rho)-ig[H,\rho]$ a classical continuous time Markov chain $(X_t)_{t\geq 0}$ on the state space 
$\{|\ffi_j^\tau\ket\bra\ffi_j^\tau |\}_{1\leq j \leq n_C}$ identified with 
$\{1,2,\dots, n\}$ with $n=n_C$, as follows.

Let us recall the general framework. The Markov process $(X_t)_{t\geq 0}$ is characterised by the probability to find the process in state $j$ at time $t\geq 0$, given the process at time $0$ is in state $i$, is denoted by
\be
p_{ij}(t)= \P(X_t=j|X_0=i), \ \ i,j\in \{1,2,\dots, n\}.
\ee
These transition probabilities $P(t)=(p_{ij}(t))_{1\leq i,j\leq n}$ are solutions to the matrix form forward and backward equations 
\be
P'(t)=P(t)Q, \ \ {P(0)=\un } \ \ \Leftrightarrow \ \ P'(t)^T=Q^TP^T(t), \ \ {P(0)=\un },
\ee
where $Q=(q_{ij})_{1\leq i,j\leq n}$ is a transition rate matrix such that $q_{ii}\leq 0$, $q_{ij}\geq 0$ and $\sum_{j=1}^nq_{ij}=0$. Hence, with the identification $Q=-{\mathfrak h}^T$ we get the following interpretation
\begin{thm}\label{thmmarkov}
Consider $\cL_g(\cdot)=\cD(\cdot)-ig[H,\cdot]$ under assumptions {\bf Spec($\overline{H}^{\, \tau}$)}
and {\bf Coup}. Then, the operator  $e^{tg^2\Phi_D}$ arising in the approximation of $e^{t\cL_g}$ provided in (\ref{appdyn}), describes a (rescaled) continuous time Markov process 
$(X_t)_{t\geq 0}$ on the state space $\{|\ffi_j^\tau\ket\bra\ffi_j^\tau |\}_{1\leq j \leq n_C}\equiv\{1,\dots, n \}$ such that for all $\tau\geq 0$,
\be\label{transphi}
\P(X_s=j|X_0=i)= \tr_C\big\{ |\ffi_i^\tau\ket\bra\ffi_i^\tau | e^{s \Phi_D}(|\ffi_j^\tau\ket\bra\ffi_j^\tau |)\big\}.
\ee 
\end{thm}
\begin{rem}
Therefore, for any $s \geq 0$, the transpose of $e^{s \Phi_D}$ is a stochastic matrix. 
\end{rem}
\aj{ Let us note that appearance of a classical Markov process on the eigenstates of the leading order driving Hamiltonian within the derivation of Lindblad generators for open quantum systems is well known. By contrast, in absence of leading order driving Hamiltonian, the state space of the Markov process into play is determined by the eigenstates of the averaged first order Hamiltonian $\overline{H}^\tau$, which takes into account the effects of the reset matrices.}\\

Finally, let us address the computation of the order $g^2$ corrections (\ref{simplenz}) of the simple eigenvalues $\lambda_{jk}(g)$ of $\cL_g(\cdot)=\cD(\cdot)-ig[H,\cdot]$ given by
\be
\tilde \lambda_{jk}^{(1)}=\tr \Big\{(\un_A\otimes|\ffi_k^\tau\ket\bra \ffi_j^\tau |\otimes \un_B)\big[H,\cL_0^{-1}([H,\tau_A\otimes|\ffi_j^\tau\ket\bra \ffi_k^\tau |\otimes \tau_B])\big]\Big\}.
\ee
We prove in Appendix that
\begin{prop} \label{prorel} Consider $\cL_g(\cdot)=\cD(\cdot)-ig[H,\cdot]$ under assumptions {\bf Spec($\overline{H}^{\, \tau}$)}
and {\bf Coup}. Then, 
the eigenvalues $ \lambda_{jk}(g)$ of $\cL_g$, see Proposition \ref{spec2}, satisfy
\be
\Re \lambda_{jk}(g) \leq -g^2\frac{\gamma_A^2+\gamma_A\gamma_B+\gamma_B^2}{\gamma_A\gamma_B(\gamma_A+\gamma_B)}(e_j^\tau-e_k^\tau)^2+\ode(g^3)
\ee
\end{prop}
\begin{rem}
Actually, we show that $\Re \tilde \lambda_{jk}^{(1)}$ is upper bounded by a sum of non positive explicit contributions. Hence one can decrease the contributions stemming from these eigenvalues in the approximations of the dynamics shown in Corollary \ref{dynimp} by assuming the coupling Hamiltonian $H$ makes the lower bounds of Lemma \ref{lemb} below  large enough.
\end{rem}

\section{Example on $\C^2\otimes \C^N\otimes\C^2$}
\label{sec:exampleN}

We present here an example where the two parts of the Hilbert space on which the dissipator acts non trivially are both $\C^2=\cH_A=\cH_B$, while the central part $\cH_C=\C^N$, with $N$ arbitrary. The orthonormal bases of $\cH_A$, $\cH_B$ and $\cH_C$ are denoted respectively by $\{|g\ket, |e\ket\}$, $\{|\downarrow\ket, |\uparrow\ket\}$ and $\{\ffi_j\}_{j=1}^N$. The reset states associated with rates $\gamma_A, \gamma_B>0$ are 
\be\label{extauab}
\tau_A=t_A |g\ket\bra g|+(1-t_A)|e\ket\bra e|, \ \  \tau_B=t_B |\downarrow\ket\bra \downarrow|+(1-t_B)|\uparrow\ket\bra \uparrow|,
\ee
where $0<t_A,t_B <1$.
We consider again a case without leading order Hamiltonian drive, that is $H_A=H_B=H_C=0$, while the order $g$ Hamiltonian reads
\begin{align}\label{exnqcq}
H&=H_\alpha\otimes \un_B+\un_A\otimes H_\beta, \ \mbox{where}\\ \nonumber
H_\alpha&=\sum_{j=1}^N a_j^{(g)}|g\otimes \ffi_j\ket \bra g\otimes \ffi_j |+ a_j^{(e)}|e\otimes \ffi_j\ket \bra e\otimes \ffi_j |+\sum_{k=1}^N\alpha_k|g\otimes \ffi_1 \ket\bra e\otimes \ffi_k | +
\mbox{h.c.}\\ \nonumber
H_\beta&=\sum_{j=1}^N b_j^{(\downarrow)}|\ffi_j\otimes\downarrow\ket \bra \ffi_j\otimes\downarrow |+ b_j^{(\uparrow)}|\ffi_j\otimes\uparrow\ket \bra \ffi_j\otimes\uparrow |+\sum_{k=1}^N\beta_k|\ffi_N\otimes\downarrow \ket\bra \ffi_k\otimes\uparrow| +
\mbox{h.c.}
\end{align}
In other words,
\begin{align}
&H_\alpha= |g\ket \bra g|\otimes H_{a}^{(g)} +  | e \ket \bra e|\otimes H_{a}^{(e)}+ |g\ket\bra e| \otimes |\ffi_1 \ket\bra \Phi_\alpha | + |e\ket\bra g|\otimes  |\Phi_\alpha \ket\bra  \ffi_1 | \\
&H_\beta= |\downarrow \ket \bra \downarrow |\otimes H_{b}^{(\downarrow)} +  | \uparrow \ket \bra \uparrow|\otimes H_{b}^{(\uparrow)}+ |\downarrow\ket\bra \uparrow| \otimes |\ffi_N \ket\bra \Phi_\beta | + |\uparrow\ket\bra \downarrow|\otimes  |\Phi_\beta \ket\bra  \ffi_N |
\end{align}
with $H_a^{(\#)}=\sum_{j=1}^N a_j^{(\#)}|\ffi_j\ket \bra \ffi_j |$, $\#\in\{g,e\}$,  $\Phi_\alpha= \sum_{k=1}^N\alpha_k\ffi_k$, and similarly for $H_\beta$,  introducing $H_b^{(\#)}=\sum_{j=1}^N b_j^{(\#)}|\ffi_j\ket \bra \ffi_j |$, $\#\in\{\downarrow, \uparrow\}$, and  $\Phi_\beta= \sum_{k=1}^N\beta_k\ffi_k$.
\medskip

\aj{On the one hand, this example shows our hypotheses can be checked for arbitrary $N$ and, on the other hand, it can lead to physically relevant models under additional assumptions, see for instance Section \ref{sec:example2} where we deal with qubits ($N=2$) subject to inter-qubit Coulomb interaction and flip-flop type interaction Hamiltonian.}
\medskip

With these definitions we compute
\begin{align}
\overline{H}^{\, \tau}=&\,  t_A H_{a}^{(g)}+ (1-t_A)H_{a}^{(e)}+ t_B H_{b}^{(\downarrow)}+ (1-t_B)H_{b}^{(\uparrow)} \nonumber \\ 
=&\sum_{j=1}^N \big(t_A a_j^{(g)}+ (1-t_A) a_j^{(e)} +t_B b_j^{(\downarrow)}+(1-t_B)b_j^{(\uparrow)} \big)|\ffi_j\ket \bra \ffi_j |,
\end{align}
which yields 
\be
\ffi_j^{\tau}=\ffi_j \ \ \mbox{ and }\ \ e_j^\tau=\big(t_A a_j^{(g)}+ (1-t_A) a_j^{(e)} +t_B b_j^{(\downarrow)}+(1-t_B)b_j^{(\uparrow)} \big). 
\ee
 We can choose the real parameters 
$a_j^{(g)}, a_j^{(e)}, b_j^{(\downarrow)}, b_j^{(\uparrow)}$  so that the generic assumption {\bf Spec $\overline{H}^{\, \tau}$} holds for any choice of $0 <t_A, t_B <1$. 

\subsection{Leading order term}
The next step consists in determining the diagonal elements of the nonnegative operators $h(k)$ defined in (\ref{posh}), $1\leq k\leq N$; more precisely $h_j(k):=\bra \ffi_j |h(k) \ffi_j\ket$, for $j\neq k$. 
We first compute
\begin{align}
\overline{H}^{\, \tau_A}&=H_\beta+(t_AH_a^{(g)}+(1-t_A)H_a^{(e)})\otimes \un_B\\
\overline{H}^{\, \tau_B}&=H_\alpha+\un_A\otimes (t_BH_b^{(\downarrow)}+(1-t_B)H_b^{(\uparrow)}).
\end{align}

Since we do not need the elements $\bra \ffi_k |h(k) \ffi_k\ket$, we do not make explicit {their contribution,} 
that we generically denote below by $c_i(k) P_k$, where $c_i(k)\geq 0$, $i=1,2,3,4$. With this convention, we get for the different elements $h(k)$ is made of
\bea
 \tr_{AB}(H(\tau_A\otimes P_k^\tau\otimes \tau_B)H)&=& c_1(k) P_k + (1-t_A)|\alpha_k|^2|\ffi_1\ket\bra\ffi_1 |+\delta_{k,1}t_A|\Phi_\alpha \ket\bra\Phi_\alpha | \nonumber \\
&& + (1-t_B)|\beta_k|^2|\ffi_N\ket\bra\ffi_N |+\delta_{k,N}t_B|\Phi_\beta \ket\bra\Phi_\beta | \nonumber \\
\tr_{A}(\overline{H}^{\, \tau_B}(\tau_A\otimes P_k^\tau )\overline{H}^{\, \tau_B}))&=&c_2(k) P_k+(1-t_A)|\alpha_k|^2|\ffi_1\ket\bra\ffi_1 |+\delta_{k,1}t_A|\Phi_\alpha \ket\bra\Phi_\alpha | \nonumber \\
\tr_{B}(\overline{H}^{\, \tau_A}( P_k^\tau \otimes \tau_B )\overline{H}^{\, \tau_A}))&=&c_3(k) P_k+(1-t_B)|\beta_k|^2|\ffi_N\ket\bra\ffi_N |+\delta_{k,N}t_B|\Phi_\beta \ket\bra\Phi_\beta |. \nonumber \\
&&
\eea

Eventually,
\bea
h(k)&=& \frac{2}{\gamma_A \gamma_B}\Big\{ (1-t_A)|\alpha_k|^2 \gamma_B  |\ffi_1\ket\bra\ffi_1 | +\delta_{k,1}t_A \gamma_B |\Phi_\alpha \ket\bra\Phi_\alpha |\\ \nonumber
&&+\delta_{k,N} t_B  \gamma_A |\Phi_\beta \ket\bra\Phi_\beta |+(1-t_B) |\beta_k|^2 \gamma_A |\ffi_N\ket\bra\ffi_N | \Big\} +c_4(k) P_k\,,
\eea

The offdiagonal elements $h_j(k)$, $j\neq k$, of the matrix $-\Phi_D$ immediately follow:  let
\begin{align}
S_k&=\gamma_B (1-t_A)|\alpha_k|^2,  \hspace{1.1cm}   U_k=\gamma_B t_A|\alpha_k|^2, &\ \ \mbox{for } \ 2\leq k\leq N \nonumber\\
T_k&=\gamma_A (1-t_B) |\beta_k|^2,  \hspace{1.1cm} V_k=\gamma_A t_B |\beta_k|^2,  &\ \ \mbox{for } \ 1\leq j \leq N-1.
\end{align}
Therefore the matrix form (\ref{matphid}) of the operator $\Phi_D$ reads
\begin{align}\label{modphi} 
\Phi_D=&\frac{-2}{\gamma_A\gamma_B}\nonumber \\
&\times \begin{pmatrix}
\tilde h_1(1) & -S_2 & -S_3 & -S_4 & \cdots & -S_{N-1}&-S_N -V_1\cr
-U_2 & \tilde h_2(2)& 0 &0&\hdots & 0 & -V_2 \cr
-U_3 & 0 & \tilde h_3(3) & 0 & \hdots& 0 &-V_3 \cr
\vdots & & \vdots & \ddots& & 0&\vdots \cr
-U_{N-1} & 0& 0& \hdots &0&\tilde h_{N-1}(N-1)&  -V_{N-1} \cr
-U_N-T_1 & -T_2 &-T_3 & -T_4 & \cdots& -T_{N-1}& \tilde h_N(N)
\end{pmatrix},
\end{align}
where the diagonal elements $\tilde h_j(j)=T_j+S_j $, $0<j<N$, $\tilde h_1(1)=\sum_{j=2}^{N} U_j+T_N$ and $\tilde h_N(N)=\sum_{j=2}^{N} V_j+S_N$.

Note that $\alpha_j\neq 0 \Leftrightarrow $ $S_j\neq 0$ and $U_j\neq 0$ , while $\beta_j\neq 0\Leftrightarrow$ $T_j\neq 0$ and $V_j\neq 0$. 
Hence, looking at the first row of (\ref{modphi}), one sees that  {\bf Coup} holds for this model when 
\be \label{hypal}
\alpha_2 \alpha_3 \dots \alpha_{N-1}\neq 0 \ \mbox{ and } \ |\beta_1|^2+|\alpha_N|^2\neq 0,
\ee
 or, looking at the last row, when 
 \be\label{hypbe}
 \beta_2\beta_3\dots\beta_{N-1}\neq 0\ \mbox{ and } \  |\beta_1|^2+|\alpha_N|^2\neq 0.
 \ee
In either cases, this validates the conclusions of Theorem \ref{astat} on the invariant state and the way to compute it. From now on, we assume that either (\ref{hypal}) or (\ref{hypbe}) holds.

The leading term of the invariant state is determined by the one dimensional kernel of $\Phi_D$ which turns out to be computable explicitly. We have, noting that $S_j+T_j>0$ for $2\leq j\leq N-1$,
\be
\ker \Phi_D=\C \begin{pmatrix} x_1 \cr x_2 \cr \vdots \cr x_N  \end{pmatrix}, \ \ \mbox{where }\ \ 
\left\{ \begin{matrix} 
x_1=S_N+\sum_{j=2}^{N-1}\frac{V_jS_j}{S_j+T_j}+V_1 \phantom{ww} \cr 
x_N=U_N+\sum_{j=2}^{N-1}\frac{U_jT_j}{S_j+T_j}+T_1 \phantom{ww}\cr 
x_j=\frac{U_jx_1+V_jx_N}{S_j+T_j}, \  2\leq j\leq N-1\end{matrix}\right. .
\ee
The corresponding faithful  leading order \aj{$\rho_0$, {\it i.e.} $\rho_0>0$,} of the invariant state of the \qrm thus reads
\be\label{leadzero}
\rho_0=\frac{1}{Z}\tau_A\otimes \sum_{j=1}^Nx_j |\ffi_j\ket\bra \ffi_j| \otimes \tau_B, \ \ \mbox{where } \ Z=\sum_{k=1}^Nx_k.
\ee
Actually, the following more explicit expressions are true. With
\be
y(N)=\sum_{j=2}^{N-1}\frac{\gamma_A\gamma_B|\alpha_j \beta_j|^2}{(1-t_A)\gamma_B|\alpha_j|^2+(1-t_B)\gamma_A|\beta_j|^2},
\ee
we can write
\begin{align}
x_1&= (1-t_A)|\alpha_N|^2\gamma_B+y(N)t_B(1-t_A)+t_B|\beta_1|^2\gamma_A   \\
x_N&= t_A|\alpha_N|^2\gamma_B+y(N)t_A(1-t_B)+(1-t_B)|\beta_1|^2\gamma_A   \\
x_j&= t_A|\alpha_N|^2\gamma_B+t_B|\beta_1|^2\gamma_A+y(N)t_At_B\\ \nonumber
&\hspace{1.2cm}+ \frac{\gamma_A\gamma_B (|\alpha_N|^2t_A(2t_B-1)|\beta_j|^2+|\beta_1|^2t_B(2t_A-1)|\alpha_j|^2)}{(1-t_A)\gamma_B|\alpha_j|^2+(1-t_B)\gamma_A|\beta_j|^2},
\end{align}
for $2\leq j \leq N$.

Note in particular the generic nontrivial dependence on $j$ of the populations of (the reduced) leading order $\rho_0$ of the invariant state.
Further remarks are in order:
\begin{itemize}
\item For non zero coefficients $\alpha_j$ and $\beta_j$, $x_j$ is independent of $2\leq j\leq N-1$ if 
\be
\frac{(2t_A-1)t_B}{(1-t_A)}\gamma_A|\beta_1|^2=\frac{(2t_B-1)t_A}{(1-t_B)}\gamma_B|\alpha_N|^2.
\ee

\item In case we consider thermal states for $\tau_\#$ on $\C^2$, $\#\in\{A,B\}$, such that $t_\#=\frac{1}{1+e^{-\beta_\#E_\#}}$, with excitation energy $E_\#>0$. We get that 
$t_\#\ra 1$ when $\beta_\#\ra \infty$, while $t_\#\ra 1/2$ when $\beta_\#\ra 0$, which shows that at high temperature, the populations tend to be constant.

\end{itemize}

\section{Example on $\C^2\otimes \C^2\otimes\C^2$}
\label{sec:example2}

With the previous example considering $H_C \in \C^N$, we could derive the exact expressions of the map $\Phi_D$ and of the leading order solution. However, going to first order correction and beyond requires considerable effort and would not be enlightening for the reader. This motivates this second example, where we restrict $H_C$ to be in $\C^2$ and consider an interaction Hamiltonian $H$ that is appropriate to describe effective physical systems. The goal of this section is twofold. First, we derive explicitly higher order corrections illustrating the theorems of Sec. \ref{sec:5}, showing that we can capture the main features of the dynamics with relatively little effort as compared to the complexity of the system. Second, we make a clear connection between a tri-partite quantum reset model and models suitable to describe realistic physical systems.

\subsection{Model}

Explicitly, we consider here a chain of three qubits characterized by their bare energies $e_A, e_B, e_C$ entering $H_0$. They are interacting through $H$. The two Hamiltonians are given by
\bea
H_0 &=& e_A |1\ket\bra 1 | \otimes \un_C \otimes \un_B + \un_A \otimes e_C |1\ket\bra 1 | \otimes \un_B + \un_A \otimes \un_C \otimes e_B |1\ket\bra 1 |\,, \\
H &=& U\, (|11\ket_{AC} \bra 11 | \otimes \un_B + \un_A \otimes (|11\ket_{CB} \bra 11 | ) \nonumber \\
&& + ( J_\alpha |01\ket_{AC} \bra 10 | \otimes \un_B + \un_A \otimes J_\beta |01\ket_{CB} \bra 10 | +  h.c)\,.
\eea
Without loss of generality, we assume the interaction strengths $U,J_\alpha, J_\beta$ to be real. This model could be effective for instance for three qubits subject to nearest-neighbour interactions: a Coulomb interaction (set by $U$) whenever two adjacent qubits are occupied and to a flip-flop interaction term of the form $\vert 01 \ket \bra 10 \vert + h.c.$ \aj{that conserves the number of excitations} (set by $J_\alpha, J_\beta$ with $J_\alpha \neq J_\beta$).
In the ordered computational basis of the three qubits 
\be\label{ordbasis}
\{ |000\ket,  |001\ket,  |010\ket,  |011\ket,  |100\ket,  |101\ket,  |110\ket,  |111\ket\},
\ee
 the total Hamiltonian $H_{tot} = H_0 + g H$ reads
\begin{align}
&H_{tot} = \label{222}\\ \nonumber
&\left( \begin{array}{cccccccc}
0 & 0& 0 & 0 & 0 & 0 & 0 & 0 \\
0 & e_B & g  J_\beta & 0 & 0 &0 & 0 & 0 \\ 
0 & g J_\beta & e_C & 0 & g J_\alpha &0 & 0 & 0 \\ 
0 & 0 & 0 & e_B + e_C +g  U & 0 &g J_\alpha & 0 & 0 \\ 
0 & 0 &g  J_\alpha & 0 & e_A &0 & 0 & 0 \\ 
0 & 0 & 0 & g J_\alpha & 0 & e_A+e_B &g  J_\beta & 0 \\ 
0 & 0 & 0 & 0 & 0 & g J_\beta & e_A + e_C + g U & 0 \\ 
0 & 0 & 0 & 0 & 0 &0 & 0 & e_A + e_B + e_C + 2 g U \\ 
\end{array}\right).
\end{align}
This model corresponds exactly to the previous example with $N=2$ and setting:
\bea\label{coefexample}
&&  \alpha_1=\beta_2=a_2^{(e)} =a_1^{(g)} =a_2^{(g)} =b_2^{(\downarrow)}= {b_2^{(\uparrow)}} =b_1^{(\uparrow)}  = 0 \\
&& a_1^{(e)} = b_1^{(\downarrow)} = U, \ \alpha_2=J_\alpha, \ \beta_1=J_\beta, 
\eea
The ground state for the three qubits is now simply given by $\vert 000 \ket$ and corresponds to {$\vert g \otimes \ffi_2 \otimes \uparrow \ket$} in the previous example with $N=2$. For clarity, we provide the expression of $H_{tot}$ in the form introduced in \eqref{exnqcq}
\bea
H_\alpha&= U |e\otimes { \ffi_1}\ket \bra e\otimes {\ffi_1} | +J_\alpha | {g}\otimes \ffi_1 \ket\bra {e}\otimes \ffi_2 |  +
\mbox{h.c.}\\ \nonumber
H_\beta&= U |{ \ffi_1 }\otimes {\downarrow} \ket \bra {\ffi_1}\otimes {\downarrow} |+ J_\beta | \ffi_2\otimes\downarrow \ket\bra \ffi_1\otimes\uparrow|  +
\mbox{h.c.}
\eea
The two ends ($A$ and $B$) of the 3-qubit chain are weakly coupled to their own thermal baths at inverse temperatures $\beta_A$ and $\beta_B$ with coupling strengths $\gamma_A$ and $\gamma_B$ respectively. Dissipation takes place following \qrm. The reset states are assumed to be thermal states defined by the Maxwell-Boltzmann distribution with their respective inverse temperature $\beta_\# = 1/T_\#$ ($k_B = 1$ in the following) in the basis $\{ \vert 0 \ket, \vert 1 \ket \}$:
\be
\tau_\#=\frac{1}{Z_\#}\left( \begin{array}{cc}
1&0 \\ 0 & e^{-\beta_\#e_\#}
\end{array}\right)=\left( \begin{array}{cc}
t_\#&0 \\ 0 & 1-t_\#
\end{array}\right), \ \ \#\in \{A, B\}.
\ee
Note that since the ground state $\vert 0 \ket$ in the $C$ part of the Hilbert space corresponds to $|\uparrow\ket$, the substitution $t_B\rightarrow (1-t_B)$ is in order to use the results of Section \ref{sec:exampleN}.

Let us remark that this model for a tri-partite open quantum system differs from previous works on reset models in the context of quantum thermodynamics, studiying in particular quantum absorption refrigerators and entanglement engines, Refs. \cite{Skrzypczyk11, Brask15, Tavakoli20}. These models consist of a chain of 2, 3 or N qubits, each of them being coupled to its own thermal bath. Dissipation due to the presence of environments is captured through \qrm. In Ref.~\cite{Skrzypczyk11}, the steady-state solution for 3 qubits with three different environments is derived analytically, whereas the case of two qubits is fully solved in Ref.~\cite{Brask15}. In contrast, in this work, we derive the steady-state solution considering an arbitrary system $C$ only coupled to the two ends $A$ and $B$ of the chain, as long as $H_C$ satisfies generic assumptions.

\subsection{Generic assumptions}

We first check the assumptions for $H_C$ and $H$.
The condition ${\bf Spec}(H_C)$, is trivially satisfied in this case as the spectrum $\sigma(H_C) = \{ 0, e_C\}$ is simple with $e_C \neq 0$. We can then verify ${\bf Spec}(\overline{H}^\tau)$ with
\be
\overline{H}^\tau = \tr_{AB}( H \tau_A \otimes \un_C \otimes \tau_B)=U(2-t_A-t_B) |1\ket\bra 1 |\,,
\ee
as defined by Eq. \eqref{hbar}\,. The spectrum $\sigma(\bar{H}^\tau) = \{ 0, U ( 2 - t_A - t_B)\}$ with associated eigenvectors $\{|0\ket, |1\ket\}$ is simple whenever $U \neq 0$ and $t_A + t_B \neq 2$ where $t_A, t_B$ stand for the ground state populations of the reset states $\tau_A, \tau_B$. The identity $t_A + t_B = 2$ is only satisfied for zero temperature reservoirs,
$t_A = t_B = 1$. 
Hence we stay in the generic case, $t_A,  t_B < 1$. The condition $U \neq 0$ also tells us that a flip-flop interaction Hamiltonian of the form $(\vert 0 1 \ket \bra 1 0 \vert + h.c.)$ is not sufficient to ensure the required non-degeneracy conditions in the 0-subspace of $\cL_0$. 
We easily verify that the kernel of $\cL_0$ has dimension $n_C^2 = 4$ if $H_C = 0$ and $n_C = 2$ if $H_C \neq 0$.

In the following, we will restrict the derivation of the steady-state solution up to the second order correction assuming no drive, {\it i.e.} $H_A = H_B = H_C = 0$. Let us note that 
in two dimensions, there is no loss of generality to consider the reset states $\tau_A$ and $\tau_B$ defined as thermal states with respect to  $H_A$ and $H_B$.

\subsection{Leading order solution, no drive}

Under {\bf Spec}($\bar{H}^\tau$) and Lemma \ref{redop}, the first-order-correction projector $\tilde{Q}_0 \tilde{\cL}_1 \tilde{Q}_0$ in the 0-eigenvalue subspace is fully characterized by the map 
$\Phi$ acting onto $\cH_C$, see Eq. \eqref{phi} and Theorem  \ref{goldenrule}
\be
\Phi(\cdot):= {\em \tr}_{AB}\big(\big[H, \cL_0^{-1}([H, \tau_A\otimes \diag(\, \cdot \,) \otimes \tau_B])\big]\big) : \cB(\cH_C)\ra \cB(\cH_C)\cap \{\rho_C\, | \, \tr \rho_C=~0\}.
\ee
In contrast to the previous example, we can compute  explicitly here the map $\Phi$ and not only $\Phi_D$. To this end, we consider $\rho_C$ to be initially in an arbitrary diagonal state (with respect to the eigenbasis of $\overline{H}^\tau$)
\bea\label{rhocdiag}
\rho_C  = \left( \begin{array}{cc}
r_C^0 & 0 \\
0 & r_C^1 \end{array} \right)\,.
\eea
Defining the linear form on $\C^2$
\be
X(r_C^0,r_C^1) =- r_C^0 (J_\beta^2 (1-t_B) \gamma_A +  J_\alpha^2 (1-t_A) \gamma_B)+ r_C^1 (J_\beta^2 t_B \gamma_A + J_\alpha^2 t_A \gamma_B)  \,,
\ee
we find the matrix $\Phi (\rho_C) \in \cB(\cH_C)$ to be given by (with respect to the eigenbasis of $\overline{H}^\tau$)
\bea
\label{eq:PhiDop}
\Phi (\rho_C) = \frac{2}{\gamma_A \gamma_B}  \left( \begin{array}{cc}
X(r_C^0,r_C^1) & 0 \\
0 &- X(r_C^0,r_C^1)
\end{array}\right)\,.
\eea 
Note that $\Phi (\rho_C)$ is diagonal, so that for this example we have $\Phi(\cdot)=\Phi_D(\cdot)$. In particular 
\be
\ker \Phi_D(\cdot)=\C \left( \begin{array}{cc}
 \gamma_A t_B J_\beta^2 + \gamma_B t_A J_\alpha^2 & 0 \\
0 & \gamma_A (1-t_B) J_\beta^2 + \gamma_B (1-t_A) J_\alpha^2 \end{array} \right)
\ee 
is one dimensional, so that Assumption {\bf Coup} is satisfied. Then $\ker \Phi_D$ provides the leading order steady-state solution
$\rho_{0} = \tau_A \otimes \rho_C^{(0)} \otimes \tau_B$
with
\bea
\rho_C^{(0)}= \left( \begin{array}{cc}
 \frac{ \gamma_A t_B J_\beta^2+  \gamma_B t_A J_\alpha^2}{\gamma_A J_\beta^2+ \gamma_B J_\alpha^2} & 0 \\
0 & \frac{\gamma_A (1-t_B) J_\beta^2+ \gamma_B (1-t_A) J_\alpha^2}{\gamma_A J_\beta^2+ \gamma_B J_\alpha^2} \end{array} \right)\,.
\eea

Interestingly, the zeroth order solution is the exact solution in the equilibrium situation, i.e. when $\tau_A = \tau_B = \tau$, the state  
{$\rho_0 = \
\tau\otimes\tau\otimes\tau$} satisfies for any $g\in\R$ (or $\C$)
$
\cL_g (\rho_0) = 0,
$
an instance of Remark ii) \ref{strucR}.

\begin{rem}
In this example, the matrix $\Phi_D$ can also be derived directly from the previous example with $N=2$, starting from the positive operator $h(k)$:

\bea
h(k) = \frac{2}{\gamma_A \gamma_B} &\Big(& \!\! \vert \alpha_k \vert ^2 (1- t_A) \gamma_B \vert 1\ket \bra 1 \vert + \vert \beta_k \vert^2 t_B \gamma_A \vert 0 \ket \bra 0 \vert \nonumber \\
 && \, + t_A \gamma_B \vert 0 \ket \bra 0 \vert + (1-t_B) \gamma_A \vert 1\ket \bra 1 \vert \,\, \Big) + c_4(k) P(k)\,.
\eea
In the basis $\{ \vert 0 \ket\bra 0|, \vert 1 \ket\bra 1|\}$, given (\ref{coefexample}), the substitution $t_B\rightarrow 1-t_B$, and  according to (\ref{modphi}),
the matrix $\Phi_D$ reads
\bea\label{matphid}
\Phi_D= \frac{-2}{\gamma_A \gamma_B}  \left( \begin{array}{cc}
\gamma_A J_\beta^2 (1-t_B) + \gamma_B J_\alpha^2 (1-t_A) &  - \gamma_A J_\beta^2 t_B - \gamma_B J_\alpha^2t_A \\
-\gamma_A J_\beta^2 (1-t_B) - \gamma_B J_\alpha^2 (1-t_A) &  \gamma_A J_\beta^2 t_B + \gamma_B J_\alpha^2 t_A 
\end{array}\right)\,,
\eea
whose kernel in this same basis is generated by the two-dimensional vector
\bea
\ker \Phi_D = \C (\gamma_A J_\beta^2 t_B + \gamma_B J_\alpha^2 t_A, \gamma_A J_\beta^2 (1-t_B) + \gamma_B J_\alpha^2 (1-t_A))^T\,.
\eea
Let us note that $\Phi_D$, when written as a superoperator acting onto diagonal matrices, takes a diagonal form, see Eq. \eqref{eq:PhiDop}. 
\end{rem}

\subsection{Underlying Markov process}
We have enough information here to determine the natural two-state classical continuous Markov process associated to the model, according to Theorem \ref{thmmarkov}. 
The state space is denoted by $\{0,1\}\equiv \{| 0 \ket\bra 0 |, | 1 \ket\bra 1 |\}$, and by (\ref{transphi}) we need to compute $e^{s \Phi_D}$ to determine the transition probabilities of the process
\be
\P(X_s=j|X_0=k)=\tr\big\{ |k\ket\bra k | e^{s \Phi_D}(| j \ket\bra j |)\big\}\equiv  (e^{s\Phi_D})_{k,j} , \ \ 0\leq j,k\leq 1
\ee
The spectral decomposition of $\Phi_D$ in the matrix form  (\ref{matphid}) is easily obtained. Introducing
\be
\ffi_+=\gamma_A J_\beta^2 t_B + \gamma_B J_\alpha^2 t_A, \ \ \ffi_-=\gamma_A J_\beta^2 (1-t_B) + \gamma_B J_\alpha^2 (1-t_A),
\ee
we have 
\be
\sigma (\Phi_D)=\{0, -2(\ffi_++\ffi_-)/(\gamma_A \gamma_B)\}=\{0, -2(\gamma_A J_\beta^2  + \gamma_B J_\alpha^2 )/(\gamma_A \gamma_B)\},
\ee
with eigenvector associated to the non zero eigenvalue proportional to $\begin{pmatrix}1 & -1 \end{pmatrix}^T$. Hence,
\be
\Phi_D=\frac{-2(\ffi_++\ffi_-)}{\gamma_A \gamma_B}Q_+ + 0\, Q_0,
\ee
with spectral projectors 
\be
Q_0=\frac{\Big| \begin{pmatrix}\ffi_+ \\ \ffi_- \end{pmatrix}\Big\rangle \Big\langle  \begin{pmatrix}1 \\ 1 \end{pmatrix} \Big|}{\ffi_-+\ffi_+} , \ \ Q_+=\frac{\Big| \begin{pmatrix}1 \\ -1 \end{pmatrix}\Big\rangle \Big\langle \begin{pmatrix}\ffi_- \\ -\ffi_+ \end{pmatrix} \Big|}{\ffi_-+\ffi_+}.
\ee
Therefore, with $\tilde s=\frac{2s(\ffi_++\ffi_-)}{\gamma_A \gamma_B}$,
\be
e^{s \Phi_D}=e^{-\tilde s}Q_+ + Q_0=\frac{1}{\ffi_-+\ffi_+} \begin{pmatrix} \ffi_+ + e^{-\tilde s}\ffi_- & \ffi_+ - e^{-\tilde s}\ffi_+\\  \ffi_- - e^{-\tilde s}\ffi_- & \ffi_- + e^{-\tilde s}\ffi_+\end{pmatrix} .
\ee
In turn this eventually yields the sought for transition probabilities
\begin{align}
\P(X_s=0|X_0=0)&=\frac{\ffi_+ + e^{-\tilde s}\ffi_-}{\ffi_-+\ffi_+}, \ \ \P(X_s=1|X_0=1)=\frac{\ffi_- + e^{-\tilde s}\ffi_+}{\ffi_-+\ffi_+},\nonumber \\
\P(X_s=1|X_0=0)&=\frac{\ffi_-(1 - e^{-\tilde s})}{\ffi_-+\ffi_+}, \ \ \P(X_s=0|X_0=1)=\frac{\ffi_+(1 - e^{-\tilde s})}{\ffi_-+\ffi_+}.
\end{align}
\aj{ We stress that in absence of leading order driving Hamiltonian, the state space of the Markov process into play is determined by the eigenstates of $\overline{H}^\tau$, that takes into account the effects of the reset matrices.}

\subsection{Higer-order corrections, no drive}

We now illustrate Theorem \ref{astat} by deriving the converging expansion of the unique invariant state of $\cL_g$
\be
\rho_0(g) = \rho_0 + g \, \rho_1 + g^2 \, \rho_2 + \ldots\,  \ \mbox{with}, \ \ 
\rho_0 = \tau_A \otimes \rho_C^{(0)} \otimes \tau_B\,,
\ee
and 
\bea
\rho_j = R_j + \tau_A \otimes r_C^{(j)} \otimes \tau_B \quad \forall j \geq 1\,. 
\eea
We recall the definitions for convenience
\bea
R_j &=& i\cL_0^{-1}([H, \rho_{j-1}]),\nonumber \\
\offdiag_\tau r_C^{(j)} &=& -i[\overline{H}^{\, \tau},  \cdot  ]^{-1}\Big( \offdiag_\tau \tr_{AB}\big(\big[H, \cL_0^{-1}([H, \rho_{j-1}])\big]\big)\Big),
\nonumber \\
\diag_\tau r_C^{(j)} &=& -\Phi_D^{-1}\big(\diag_\tau \tr_{AB}([H,\cL_0^{-1}(i[H, R_{j}+\tau_A\otimes\offdiag_\tau r_C^{(j)}\otimes\tau_B])])\big)\,. \nonumber
\eea
For the first-order correction, we start computing
$R_1=i\cL_0^{-1}([H, \tau_A \otimes \rho_C^{(0)} \otimes \tau_B])$ which can be expressed with $F_1=i(|01\ket \bra 10 | -|10\ket \bra 01 | )$ (acting on  $\cH_A\otimes\cH_C$ or $\cH_C\otimes\cH_B$ depending on the context) as
\begin{align}\label{calcR1}
R_1
=&\frac{(t_A-t_B)J_\alpha J_\beta}{\gamma_A J_\beta^2+ \gamma_B J_\alpha^2}\Big(J_\beta F\otimes\tau_B+ J_\alpha \tau_A\otimes F \Big) \nonumber \\
=&\frac{(t_A-t_B)J_\alpha J_\beta}{\gamma_A J_\beta^2+ \gamma_B J_\alpha^2}\Big( J_\beta\, i(|01\ket \bra 10 | -|10\ket \bra 01 | )\otimes\tau_B+ J_\alpha \tau_A\otimes\, i(|01\ket \bra 10 | -|10\ket \bra 01 | ) \Big).
\end{align}

We first note that since $\Phi=\Phi_D$, the expression 
for $\offdiag_\tau r_C^{(1)}$ reduces to zero:
\begin{align}\label{zerooffdiag1}
\offdiag_\tau r_C^{(1)}
&=-i[\overline{H}^{\, \tau},  \cdot  ]^{-1}\Big( \offdiag_\tau \Phi (\rho_C^{(0)})\Big)\equiv 0.
\end{align}

Then, it remains to determine $\diag_\tau r_C^{(1)}$ 
to get the first order correction in $g$. Thanks to (\ref{zerooffdiag1}) and using (\ref{calcR1}) for $R_1$, we compute
\be
\diag_\tau r_C^{(1)}=-\Phi_D^{-1}\big(\diag_\tau \tr_{AB}([H,\cL_0^{-1}(i[H, R_{1}])])\big)=0.
\ee

Hence, the first order correction is simply given by $R_1$, $\rho_1 = R_1$ and we obtain
\bea
\rho_0(g) = \tau_A \otimes \rho_C^{(0)}\otimes  \tau_B + g R_1+ \ode(g^2)\,.
\eea

We proceed with the second-order correction and compute $R_2 = i \cL_0^{-1} \big(  [H, R_1] \big)$. The matrix $R_2$ is rather complex and we provide the expressions for its diagonal and off-diagonal elements separately. Its 8 diagonal elements in the ordered basis (\ref{ordbasis}) are proportional to by
\bea \diag (R_2) &=& \frac{J_\alpha J_\beta}{\gamma_A \gamma_B} \Bigg(
t_A t_B (\gamma_A - \gamma_B), - t_A (t_B \gamma_A + \gamma_B (1-t_B)), \nonumber \\
&& t_A \gamma_A (1-t_B) - t_B \gamma_B (1 -t_A), (1-t_B) (t_A \gamma_A - \gamma_B (1-t_A)), \nonumber \\
&& t_B (\gamma_B t_A + \gamma_A (1-t_A)), \gamma_A t_A (1-t_B) - \gamma_A t_B (1-t_A),\nonumber \\
&& (1-t_A) (\gamma_B t_B + \gamma_A (1-t_B)), (1-t_A) (1-t_B) (\gamma_B - \gamma_A)\,
\Bigg).
\eea
For its off-diagonal elements, we introduce $F_2 = |01\ket \bra 10 |  + |10\ket \bra 01 \vert$ and the coefficient matrices
\bea
\Gamma_A = \left( \begin{array}{cc}
\gamma_A & 0 \\
0 & \gamma_A + \gamma_B/(1-t_A) \end{array} \right) \quad ; \quad \Gamma_B = \left( \begin{array}{cc}
\gamma_B & 0 \\
0 & \gamma_B + \gamma_A/(1-t_B) \end{array} \right)\,.
\eea
The matrix $R_2$ can then be written in a compact form
\bea
R_2 &=&  \frac{2 J_\alpha J_\beta (t_A - t_B)}{J_\beta^2 \gamma_A + J_\alpha^2 \gamma_B}  
\Big \{ \diag(R_2) + \frac{1}{\gamma_A + \gamma_B}\\ \nonumber 
&&\times\Big( \frac{- J_\alpha U (1-t_A)}{2 \gamma_B}  \tau_A \Gamma_A \otimes F_2 +  \frac{J_\beta U (1-t_B)}{2 \gamma_A} F_2 \otimes \tau_B \Gamma_B \\ \nonumber 
&&\phantom{\times} - \frac{1}{2} (J_\alpha^2 t_A - J_\beta^2 t_B) \vert 001 \ket \bra 100 \vert + \frac{1}{2} (J_\alpha^2 (1-t_A ) - J_\beta^2 (1-t_B)) \vert 110 \ket \bra 011 \vert \Big) \Big\}\,.
\eea
{For $\offdiag r_C^{(2)}$, we find that it is equal to 0. This leads us to:} 
\bea
\diag \, r_C^{(2)} = 
\left( \begin{array}{cc}
X^{(2)}& 0 \\
0 & - X^{(2)} \end{array} \right)
\eea

with
\bea
X^{(2)} &=& \frac{2 i J_\alpha^2 J_\beta^2 (t_A - t_B)}{ \gamma_A^2 \gamma_B^2 (\gamma_A + \gamma_B) (J_\beta^2 \gamma_A + J_\alpha^2 \gamma_B)}  \\
&&\times \Big\{  (\gamma_A + \gamma_B) ( J_\beta^2 \gamma_A (2 \gamma_A - \gamma_B) - J_\alpha^2 \gamma_B (2 \gamma_B - \gamma_A) )\nonumber \\  \nonumber
&&\phantom{\times} + U^2  ( (1-t_A) \gamma_A^2( \gamma_B + (1-t_A) \gamma_A) - (1-t_B) \gamma_B^2 (\gamma_A - (1-t_B) \gamma_B)) \Big\}\,.
\eea
{The solution up to the second-order correction is then given by
\bea
\rho_0(g) = \tau_A \otimes ( \rho_C^{(0)} + g^2 r_C^{(2)}) \otimes \tau_B + g \, R_1 + g^2 R_2 + \ode (g^3)\,. 
\eea
}
We note that coulomb-interaction term like in $U$ starts playing a role when considering the second-order correction.

\section{Appendix}

We provide here the proof of Proposition \ref{prorel}.

\begin{proof}
By computations similar to those performed in the determination of $\Phi_D$, we have with $P_{jk}^\tau=|\ffi_j^\tau\ket\bra \ffi_k^\tau |$,
\begin{align}\label{firstep}
[H,\cL_0^{-1}( &[H,\tau_A\otimes P_{jk}^\tau\otimes \tau_B])]=-\frac{1}{\gamma_A+\gamma_B} [H,[H,\tau_A\otimes P_{jk}^\tau\otimes \tau_B]]
\\ \nonumber
&-\frac{\gamma_B/\gamma_A}{\gamma_A+\gamma_B}[H,[\overline{H}^{\, \tau_B},\tau_A\otimes P_{jk}^\tau]\otimes \tau_B]-\frac{\gamma_A/\gamma_B}{\gamma_A+\gamma_B}[H,\tau_A\otimes [\overline{H}^{\, \tau_A}, P_{jk}^\tau\otimes\tau_B]]\\ \nonumber
&-\frac{\gamma_A^2+\gamma_A\gamma_B+\gamma_B^2}{\gamma_A\gamma_B(\gamma_A+\gamma_B)}(e_j^\tau-e_k^\tau)[H,\tau_A\otimes P_{jk}^\tau\otimes \tau_B].
\end{align}
The last term in (\ref{firstep}) yields the following contribution to $\tilde \lambda_{jk}^{(1)}$, using cyclicity of the trace and $\overline{H}^{\, \tau}\ffi_j^\tau=e_j^\tau \ffi_j^\tau$,
\be\label{newterm}
-\frac{\gamma_A^2+\gamma_A\gamma_B+\gamma_B^2}{\gamma_A\gamma_B(\gamma_A+\gamma_B)}(e_j^\tau-e_k^\tau)^2<0.
\ee
Then, using cyclicity of the trace and $P_{kj}^\tau P_{jk}^\tau=P_{kk}^\tau$, we have 
\begin{align}\label{ttt}
\tr (\un_A\otimes P_{kj}^\tau&\otimes \un_B[H,[H,\tau_A\otimes P_{jk}^\tau\otimes \tau_B]])=-2\tr(\un_A\otimes P_{kj}^\tau\otimes \un_B H \tau_A\otimes P_{jk}^\tau\otimes \tau_B H)  \nonumber  \\ \nonumber
&\hspace{2.7cm}+\tr (H(\tau_A\otimes P_{jj}^\tau\otimes \tau_B)H)+\tr (H(\tau_A\otimes P_{kk}^\tau\otimes \tau_B)H) \\
&\hspace{-2.cm}=\tr (H(\tau_A\otimes P_{jj}^\tau\otimes \tau_B)H)+\tr (H(\tau_A\otimes P_{kk}^\tau\otimes \tau_B)H)-2(\tr_{AB}(H (\tau_A\otimes P_{jk}^\tau\otimes \tau_B) H))_{jk}.
\end{align}
Here $A_{jk}$ denotes the $jk$ of the matrix $A\in\cB(\cH_C)$ with respect to the basis $\{\ffi_j^\tau\}$.
Note that the operators in the full traces  are non negative, whereas the last term is {\it a priori} complex valued.

Similarly,
\begin{align}
&\tr (\un_A\otimes P_{kj}^\tau\otimes \un_B [H,\tau_A\otimes [\overline{H}^{\, \tau_A}, P_{jk}^\tau\otimes\tau_B]])=\\ \nonumber
&\tr_{AC} (\overline{H}^{\, \tau_A}(P_{jj}^\tau\otimes \tau_B)\overline{H}^{\, \tau_A})+\tr_{AC} (\overline{H}^{\, \tau_A}(P_{kk}^\tau\otimes \tau_B)\overline{H}^{\, \tau_A})
-2(\tr_{B}(\overline{H}^{\, \tau_A}  P_{jk}^\tau\otimes \tau_B \overline{H}^{\, \tau_A}))_{jk},
\end{align}
and the analogous formula holds for the term involving $\overline{H}^{\, \tau_B}$. These expressions allow us to bound below their real part by a non negative quantity, as the next lemma shows.
\begin{lem} \label{lemb} Under the hypotheses above, we compute
\begin{align}
\Re\,\tr& \Big\{(\un_A\otimes P_{kj}^\tau\otimes \un_B)\big[H,[H,\tau_A\otimes P_{jk}^\tau\otimes \tau_B]\big]\Big\} \nonumber\\ 
&\geq \tr \Big\{(\un_A\otimes(\un_C- P_{jj}^\tau)\otimes \un_B)H(\tau_A\otimes P_{jj}^\tau \otimes \tau_B)H(\un_A\otimes(\un_C-P_{jj}^\tau )\otimes \un_B) \Big\}  \nonumber \\
&\phantom{\geq }+ \mbox{\rm same with} \ k\leftrightarrow j.\\
\Re\,\tr& \Big\{\un_A\otimes P_{kj}^\tau\otimes \un_B [H,\tau_A\otimes [\overline{H}^{\, \tau_A}, P_{jk}^\tau\otimes\tau_B]]\Big\} \nonumber\\ 
&\geq \tr \Big\{((\un_C-P_{jj}^\tau)\otimes \un_B) \overline{H}^{\, \tau_A}(P_{jj}^\tau\otimes \tau_B)\overline{H}^{\, \tau_A}((\un_C-P_{jj}^\tau )\otimes \un_B) \Big\}  \nonumber \\
&\phantom{\geq }+ \mbox{\rm same with} \ k\leftrightarrow j.\\
\Re\,\tr& \Big\{\un_A\otimes P_{kj}^\tau \otimes \un_B)\big[H,[\overline{H}^{\, \tau_B},\tau_A\otimes P_{jk}^\tau]\otimes \tau_B\big]\Big\} \nonumber\\ 
&\geq \tr \Big\{(\un_A\otimes(\un_C-P_{jj}^\tau))\overline{H}^{\, \tau_B}(\tau_A\otimes P_{jj}^\tau)\overline{H}^{\, \tau_B}(\un_A\otimes(\un_C-P_{jj}^\tau)) \Big\}  \nonumber \\
&\phantom{\geq }+ \mbox{\rm same with} \ k\leftrightarrow j.
\end{align}
\end{lem}
\begin{rem}
Since 
\begin{align}
&(\un_A\otimes(\un_C- P_{jj}^\tau)\otimes \un_B)H(\tau_A\otimes P_{jj}^\tau \otimes \tau_B)H(\un_A\otimes(\un_C-P_{jj}^\tau )\otimes \un_B)=\\ \nonumber
&((\tau_A^{1/2}\otimes P_{jj}^\tau \otimes \tau_B^{1/2})H(\un_A\otimes(\un_C-P_{jj}^\tau )\otimes \un_B))^*\\ \nonumber
&\hspace{4cm}\times((\tau_A^{1/2}\otimes P_{jj}^\tau \otimes \tau_B^{1/2})H(\un_A\otimes(\un_C-P_{jj}^\tau )\otimes \un_B))
\end{align}
is a non negative operator, we get from (\ref{firstep}), (\ref{newterm}) and the Lemma that 
\be
\Re \tilde \lambda_{jk}^{(1)}\leq -\frac{\gamma_A^2+\gamma_A\gamma_B+\gamma_B^2}{\gamma_A\gamma_B(\gamma_A+\gamma_B)}(e_j^\tau-e_k^\tau)^2<0,
\ee 
which proves Proposition \ref{prorel}.
\end{rem}
\begin{proof}
We prove the first inequality, the others are similar. Let $G=H(\tau_A^{1/2}\otimes \un_C \otimes \tau_B^{1/2})$, so that the real part we need to consider reads, see (\ref{ttt}),
\be\label{tbc}
\tr (G(\un_A\otimes P_{jj}^\tau\otimes \un_B)G^*)+\tr (G(\un_A\otimes P_{kk}^\tau\otimes \un_B)G^*)-2\tr((\un_A\otimes P_{jk}^\tau\otimes \un_B)G(\un_A\otimes P_{kj}^\tau\otimes \un_B)G^*).
\ee
Spelling out the traces we get
\begin{align}
&\tr (G(\un_A\otimes P_{jj}^\tau\otimes \un_B)G^*)=\sum_{n,m,r,s}\sum_l \big|\bra \ffi_r^{A}\otimes \ffi_l^\tau\otimes \ffi_s^{B}|G \, \ffi_n^{A}\otimes \ffi_j^\tau\otimes \ffi_m^{B} \ket \big|^2\\ \label{gjk}
&\tr ((\un_A\otimes P_{jk}^\tau\otimes \un_B)G(\un_A\otimes P_{kj}^\tau\otimes \un_B)G^*)   \\ \nonumber
&\hspace{1cm}= \sum_{n,m,r,s} \bra \ffi_r^{A}\otimes \ffi_k^\tau\otimes \ffi_s^{B}|G \, \ffi_n^{A}\otimes \ffi_k^\tau\otimes \ffi_m^{B} \ket 
 \bra \ffi_n^{A}\otimes \ffi_j^\tau\otimes \ffi_m^{B}|G \, \ffi_r^{A}\otimes \ffi_j^\tau\otimes \ffi_s^{B} \ket,
\end{align}
and we observe that the complex conjugate of (\ref{gjk}) is obtained by exchanging $j$ and $k$. 
Hence we can express the real part of (\ref{tbc}) as
\begin{align}
\sum_{n,m,r,s} &
 \big|\bra \ffi_r^{A}\otimes \ffi_j^\tau\otimes \ffi_s^{B}|G \, \ffi_n^{A}\otimes \ffi_j^\tau\otimes \ffi_m^{B} \ket \big|^2
+\big|\bra \ffi_r^{A}\otimes \ffi_k^\tau\otimes \ffi_s^{B}|G \, \ffi_n^{A}\otimes \ffi_k^\tau\otimes \ffi_m^{B} \ket \big|^2\nonumber\\
&-  \bra \ffi_r^{A}\otimes \ffi_k^\tau\otimes \ffi_s^{B}|G \, \ffi_n^{A}\otimes \ffi_k^\tau\otimes \ffi_m^{B} \ket 
 \bra \ffi_n^{A}\otimes \ffi_j^\tau\otimes \ffi_m^{B}|G \, \ffi_r^{A}\otimes \ffi_j^\tau\otimes \ffi_s^{B} \ket \nonumber\\
&-  \bra \ffi_r^{A}\otimes \ffi_j^\tau\otimes \ffi_s^{B}|G \, \ffi_n^{A}\otimes \ffi_j^\tau\otimes \ffi_m^{B} \ket 
 \bra \ffi_n^{A}\otimes \ffi_k^\tau\otimes \ffi_m^{B}|G \, \ffi_r^{A}\otimes \ffi_k^\tau\otimes \ffi_s^{B} \ket \nonumber\\
 &+\big|\bra \ffi_r^{A}\otimes \ffi_k^\tau\otimes \ffi_s^{B}|G \, \ffi_n^{A}\otimes \ffi_j^\tau\otimes \ffi_m^{B} \ket \big|^2
+\big|\bra \ffi_r^{A}\otimes \ffi_j^\tau\otimes \ffi_s^{B}|G \, \ffi_n^{A}\otimes \ffi_k^\tau\otimes \ffi_m^{B} \ket \big|^2\nonumber\\
+\sum_{n,m,r,s} &\sum_{l\not\in\{j,k\}} 
\big|\bra \ffi_r^{A}\otimes \ffi_l^\tau\otimes \ffi_s^{B}|G \, \ffi_n^{A}\otimes \ffi_j^\tau\otimes \ffi_m^{B} \ket \big|^2+
\big|\bra \ffi_r^{A}\otimes \ffi_l^\tau\otimes \ffi_s^{B}|G \, \ffi_n^{A}\otimes \ffi_k^\tau\otimes \ffi_m^{B} \ket \big|^2.
\end{align}
With $a= \bra \ffi_r^{A}\otimes \ffi_j^\tau\otimes \ffi_s^{B}|G \, \ffi_n^{A}\otimes \ffi_j^\tau\otimes \ffi_m^{B} \ket $ and $b=\bra \ffi_r^{A}\otimes \ffi_k^\tau\otimes \ffi_s^{B}|G \, \ffi_n^{A}\otimes \ffi_k^\tau\otimes \ffi_m^{B} \ket $, we rewrite the first four terms of the summand as
\be
|a|^2+|b|^2-b\bar a -a\bar b=\Big\bra\begin{pmatrix}a \cr b\end{pmatrix}\Big| \begin{pmatrix}1 & -1 \cr -1 & 1\end{pmatrix}\begin{pmatrix}a \cr b\end{pmatrix}\Big\ket \geq 0,
\ee
since $\begin{pmatrix}1 & -1 \cr -1 & 1\end{pmatrix}\geq 0$. The remaining terms can reorganised as follows,
\begin{align}
&\sum_{n,m,r,s} \big|\bra \ffi_r^{A}\otimes \ffi_j^\tau\otimes \ffi_s^{B}|G \, \ffi_n^{A}\otimes \ffi_k^\tau\otimes \ffi_m^{B} \ket \big|^2+\sum_{l\not\in\{j,k\}} 
\big|\bra \ffi_r^{A}\otimes \ffi_l^\tau\otimes \ffi_s^{B}|G \, \ffi_n^{A}\otimes \ffi_j^\tau\otimes \ffi_m^{B} \ket \big|^2\nonumber \\ 
&=\sum_{n,m,r,s} \sum_{l\neq j} 
\big|\bra \ffi_r^{A}\otimes \ffi_l^\tau\otimes \ffi_s^{B}|G \, \ffi_n^{A}\otimes \ffi_j^\tau\otimes \ffi_m^{B} \ket \big|^2\nonumber \\ 
&=
 \tr \Big\{(\un - \un_A\otimes P_{jj}^\tau\otimes  \un_B) G(\un_A\otimes P_{jj}^\tau \otimes \un_B)G^*(\un -\un_A\otimes P_{jj}^\tau \otimes \un_B) \Big\} ,
\end{align}
and similarly for the terms with second index equal to $k$, which yields the result.  \hfill $\Box$
\end{proof}
And the proof of Proposition  \ref{prorel} is finished.
\end{proof}

\medskip

\noindent {\bf Acknowledgments}: GH acknowledges the Swiss National Science Foundation through the starting grant PRIMA PR00P2$\_$179748. AJ is partially supported by the Agence Nationale de la Recherche through the grant NONSTOPS (ANR-17-CE40-0006-01), and he  wishes to thank the Universit\'e de Gen\`eve for hospitality during the first stages of this work. Both authors acknowledge support from the Banff International Research Station which hosted the 2019 meeting "Charge and Energy Transfer Processes: Open Problems in Open Quantum Systems" where this project started.


\begin{thebibliography}{99}

\bibitem{AttalPautrat} S. Attal and Y. Pautrat. From repeated to continuous quantum interactions. Ann. Henri Poincar\'e, {\bf 7}, 59-104, (2006).

\bibitem{Barra15} F. Barra, The thermodynamic cost of driving quantum systems by their boundaries. Sci. Rep. {\bf 5}, 14873, (2015).

\bibitem{Barra17} F. Barra, C. Lled\'o, Stochastic thermodynamics of quantum maps with and without equilibrium. Phys. Rev. E {\bf 96}, 052114, (2017).

\bibitem{Brask15} J. Bohr Brask, G. Haack, N. Brunner, M. Huber, Autonomous quantum thermal machine for generating steady-state entanglement, New J Phys. {\bf 17}, 113029, (2015).


\bibitem{Breuer} H. P. Breuer and F. Petruccione, The Theory of Open Quantum Systems vol 1 (Oxford
University Press), (2007).

%\bibitem{Brun02} T. A. Brun, A simple model of quantum trajectories, Am. J. Phys. {\bf 70}, 719, (2002).

\aj{ \bibitem{Brunner14} N. Brunner, M. Huber, N. Linden, S. Popescu, R. Silva, and P. Skrzypczyk, Entanglement enhances cooling in microscopic quantum refrigerators, Phys. Rev. E {\bf 89}, 032115 (2014).}

\bibitem{BJM0} L. Bruneau, A. Joye, and M. Merkli. Asymptotics of repeated interaction quantum systems. J. Funct.
Anal., {\bf 239}, 310-344, (2006).

\bibitem{BJMRev} L. Bruneau, A. Joye, and M. Merkli. Repeated interactions in open quantum systems. J. Math.
Phys., {\bf 55}, 075204, (2014).

\bibitem{BruPil}  L. Bruneau and C.-A. Pillet. Thermal relaxation of a QED cavity. J. Stat. Phys., {\bf 134}, 1071-1095, (2009).

%\bibitem{Caves87} C. M. Caves and J. G. Milburn, Quantum-mechanical model for continuous position measurements, Phys. Rev. A {\bf 36}, 5543, (1987).

\bibitem{DerFru} J. Derezi\'nski, R. Fr\"uboes, Fermi Golden Rule and Open Quantum Systems, in Open Quantum Systems III, S. Attal., A. Joye, C.-A. Pillet (Eds.), Lecture Notes in Mathematics 1882, (2006). 

\aj{ \bibitem{Evans11} M. R. Evans and S. N. Majumdar, Diffusion with Stochastic Resetting, Phys. Rev. Lett. {\bf 106}, 160601 (2011).}

%\bibitem{Gisin92} N. Gisin, I. C. Percival, The quantum-state diffusion model applied to open systems, J. Phys. A {\bf 25}, 5677-5691, (1992).

\bibitem{GKLS} V. Gorini, A. Kossakowski, and E. C. G. Sudarshan, Completely positive dynamical semigroups
of N-level systems, J. Math. Phys. {\bf17}, 821, (1976); G. Lindblad, On the generators of quantum dynamical semigroups, Commun. Math. Phys. {\bf48}, 119, (1976).

\bibitem{HJPR} E. P. Hanson, A. Joye, Y. Pautrat, and R. Raqu\'epas, Landauer's principle in repeated interaction systems, Comm. Math. Phys. {\bf 349},285-327, (2017).

\aj{\bibitem{Hartmann17} L. Hartmann, W. D\"ur and H. J. Briegel, Entanglement and its dynamics in open,
dissipative systems, New Journal of Physics {\bf 9}, 230 (2007).}

\bibitem{Kampen} N. G. van Kampen, Stochastic Processes in Physics and Chemistry (Elsevier, Amsterdam, 2007).

\bibitem{K} T. Kato, Perturbation Theory for Linear Operators (Springer-Verlag Berlin Heidelberg New York 1980).

\bibitem{KumMaa} B. K\"ummerer, H. Maassen, A scattering theory for Markov Chains, Infinite Dimensional Analysis, Quantum Probability and Related Topics, {\bf 03}, 161-176, (2000).

\aj{\bibitem{LPS} N. Linden, S. Pospescu, P. Skrzypczyk, How small can thermal machines be? The smallest possible refrigerator, Phys. Rev. Lett. {\bf 105}, 130401, (2010).}

\bibitem{Lorenzo17} S. Lorenzo, F. Ciccarello, G. M. Palma, Composite quantum collision models. Phys. Rev. A {\bf 96}, 032107, (2017).

\aj{\bibitem{KM} K. Macieszczak, M. Guta, I. Lesanovsky, J.P. Garrahan, Towards a theory of metastability in open quantum dynamics, Phys. Rev. Lett. {\bf 116}, 240404, (2016).}

\aj{\bibitem{MSM18} B. Mukherjee, K. Sengupta, Satya N. Majumdar, Quantum dynamics with stochastic reset, Phys. Rev. B {\bf 98}, 104309 (2018).}

\bibitem{N} Norris, J.R., Markov Chains (Cambridge University Press, 1997).

\bibitem{Pezzuto16} M. Pezzutto, M. Paternostro, Y. Omar, Implications of non-Markovian quantum dynamics for the Landauer bound, New J. Phys. {\bf 18}, 123018, (2016).

\bibitem{Rau63} J. Rau, Relaxation Phenomena in Spin and Harmonic Oscillator Systems, Phys. Rev. {\bf 129}, 1880, (1963).

\bibitem{Rose18} D. C. Rose, H. Touchette, I. Lesanovsky, J. P. Garrahan, Spectral properties of simple classical and quantum reset processes Phys. Rev. E {\bf 98}, 022129, (2018).

%\bibitem{Scarani02} V. Scarani M. Ziman, P. Stelmachovic, N. Gisin, and V. Buzek, Thermalizing Quantum Machines: Dissipation and Entanglement, Phys. Rev. Lett. {\bf 88}, 097905, (2002).

\bibitem{Schaller} G. Schaller, Open Quantum Systems Far from Equilibrium (Springer, Cham, 2014).

\bibitem{Seah19} S. Seah, S. Nimmrichter, V. Scarani, Nonequilibrium dynamics with finite-time repeated interactions. Phys. Rev. E {\bf 99}, 042103, (2019).

\bibitem{Skrzypczyk11} P. Skrzypczyk, N. Brunner, N. Linden and S. Popescu,  The smallest refrigerators can reach maximal efficiency, J. Phys. A: Mathematical and Theoretical {\bf 44}, 49, (2011).

\bibitem{Strasberg17} P. Strasberg, G. Schaller, T. Brandes, M. Esposito, Quantum and Information Thermodynamics: A Unifying Framework Based on Repeated Interactions, Phys. Rev. X {\bf 7}, 021003 (2017).

\bibitem{Tavakoli20} A. Tavakoli, G. Haack, N. Brunner, J. Bohr Brask, Autonomous multipartite entanglement engines, Phys. Rev. A {\bf 101}, 012315, (2020).

\bibitem{Tavakoli18} A. Tavakoli, G. Haack, M. Huber, N. Brunner, J. Bohr Brask, Heralded generation of maximal entanglement in any dimension via incoherent coupling to thermal baths, Quantum 2, 73, (2018).







\end{thebibliography}
\end{document}